\documentclass[11pt, letterpaper]{article}

\usepackage[left=1in, right=1in, top=1in, bottom=1in]{geometry}
\usepackage[utf8]{inputenc}
\usepackage[T1]{fontenc}
\usepackage{bm}
\usepackage{type1cm}
\usepackage{lettrine}
\usepackage{amsmath,amssymb,amsthm}
\usepackage{moreverb}
\usepackage{mathtools}
\usepackage{amsmath}
\usepackage{amssymb}
\usepackage{algorithmic}
\usepackage{graphics}
\usepackage{graphicx}
\usepackage{subfigure}
\usepackage{caption}
\usepackage{extarrows}
\usepackage{color}
\usepackage{framed}
\usepackage{wrapfig}
\usepackage{bm}
\usepackage{mathrsfs}
\usepackage{multirow}
\usepackage{longtable}
\usepackage{hyperref}
\usepackage{paralist}
\usepackage{indentfirst}
\usepackage{relsize}
\usepackage{extarrows}
\usepackage{upgreek}
\usepackage{bm}
\usepackage{mwe}
\usepackage[dvipsnames,table]{xcolor}

\definecolor{codedefine}{rgb}{0.15,0.35,0.55}
\definecolor{codefunc}{rgb}{0.45,0.20,0.55}
\definecolor{codecall}{rgb}{0.10,0.45,0.40}
\definecolor{codeblue}{rgb}{0.25,0.5,0.5}
\usepackage{booktabs}
\usepackage{authblk}
\usepackage{lettrine}
\usepackage{type1cm}
\usepackage{threeparttable}
\usepackage[sort&compress,numbers]{natbib}
\usepackage[figurename=Figure]{caption}
\usepackage[normalem]{ulem}
\usepackage{placeins}
\usepackage{xspace}

\graphicspath{{./Figure/}{./Figure_supp/}{./figure/}}
\usepackage[font=footnotesize,labelfont=bf]{caption}

\usepackage[percent]{overpic}
\newcommand{\method}{scEGFlow\xspace}

\newcommand{\ie}{\textit{i}.\textit{e}.}
\newcommand{\eg}{\textit{e}.\textit{g}.}

\usepackage{algorithm}
\usepackage{listings}
\usepackage{float}
\usepackage{array}
\hypersetup{
bookmarks=true,
bookmarksopen=true,
bookmarksnumbered=true,
unicode=false,
pdftoolbar=true,
pdfmenubar=true,
pdffitwindow=false,
pdfstartview={FitH},
pdftitle={Generalizable single-cell perturbation response prediction using energy-guided flow matching},
pdfauthor={Jianan Wei, Jiajun Hong, Guikun Chen, Ning Yang, Lifeng Fan, Wenguan Wang},
pdfsubject={arXiv preprint},
pdfcreator={Creator},
pdfproducer={Producer},
pdfkeywords={single-cell perturbation, flow matching, energy guidance},
pdfnewwindow=true,
colorlinks=true,
linkcolor=blue,
citecolor=blue, 
filecolor=blue,
urlcolor=blue
}

\begin{document}
\title{\textbf{Generalizable single-cell perturbation response prediction using energy-guided flow matching}}

\author[1,$^*$]{Jianan Wei}
\author[1,3,$^*$]{Jiajun Hong}
\author[1]{Guikun Chen}
\author[2]{Ning Yang}
\author[3]{Lifeng Fan}
\author[1,\dag]{Wenguan Wang}

\affil[1]{\small Zhejiang University, Hangzhou, Zhejiang, China}

\affil[2]{\small Peking University Chengdu Academy for Advanced Interdisciplinary Biotechnologies, Chengdu, Sichuan, China}

\affil[3]{\small Beijing Institute for General Artificial Intelligence, Beijing, China}
\vspace{18pt}

\date{}
\renewcommand{\thefootnote}{}
\footnotetext[1]{*\hspace{0.2em}These authors contributed equally to this work.}
\footnotetext[2]{\dag\hspace{0.2em}Corresponding author: Wenguan Wang.}
\renewcommand{\thefootnote}{\svthefootnote}

\maketitle

\normalsize

\vspace{-18pt} 
\begin{abstract}
	\small

Predicting phenotypic and transcriptional responses to perturbations at single-cell resolution provides a powerful tool for probing biological systems.
However, existing methods typically rely on fixed mappings learned during training, making it challenging to calibrate distribution shifts or adapt to novel perturbation conditions during inference.
Here, we present \method, an energy-guided flow matching framework that dynamically bridges control and perturbed cellular states.
\method models continuous transitions from control cell populations to perturbed states using conditional flow matching.
It then applies condition-specific energy gradients to correct and steer these predictions, enabling flexible adjustments without retraining the flow model.
Evaluations across benchmarks spanning imaging phenotypes and transcriptomic profiles show that \method outperforms existing methods in reconstructing response distributions under both seen and unseen perturbation conditions, faithfully preserving cellular manifold geometry and population heterogeneity. 
This advantage is notable when adapting to new conditions with only a few measured cells, consistently improving prediction accuracy.
Furthermore, \method accurately recapitulates perturbation-induced up- and down-regulation patterns across consensus gene expression signatures, where energy guidance improves the agreement between predicted and observed regulatory directions.
Ultimately, these findings  demonstrate that \method provides a modular, generalizable, and steerable solution for single-cell perturbation modeling.
By grounding generative flows in learned biological landscapes, this architecture establishes a new computational paradigm for navigating and manipulating cellular behavior in silico.
\end{abstract}


\vspace{12pt} 
\section{Introduction}

Single-cell perturbation technology provides an efficient tool to investigate complex and heterogeneous systems by profiling cellular responses to genetic, chemical, or environmental perturbations~\citep{qi2024prnet, norman2019exploring}.
However, high-throughput perturbation screens face substantial challenges due to the complexity of biological systems and the vast combinatorial space of possible perturbations, of which only a small fraction can be experimentally measured~\citep{norman2019exploring, srivatsan2020massively, replogle2022mapping}.
To circumvent this combinatorial sparsity, diverse computational methods have been developed to learn from experimental cellular profiles collected under distinct perturbation conditions and generalize to unseen perturbations.
Exemplified by scGen \citep{lotfollahi2019scgen} and CPA \citep{lotfollahi2023cpa}, latent-variable approaches leverage vector arithmetic in latent space to predict out-of-distribution cellular responses.
Another line of work improves extrapolation by incorporating biological priors. 
chemCPA~\citep{hetzel2022chemcpa} replaces a fixed drug embedding dictionary with a
molecular structure encoder, thereby extending predictions to unseen
compounds, while GEARS~\citep{roohani2024gears} exploits prior knowledge of
gene-gene relationships to model unseen genetic 
perturbations.
From a data-driven perspective, foundation models, including scGPT~\citep{cui2024scgpt} and scFoundation~\citep{hao2024large}, leverage large-scale pretraining to obtain transferable representations for downstream perturbation prediction.
Conditional generators, such as PRNet~\citep{qi2024prnet}, learn to generate post-perturbation transcriptional profiles from cellular states and  perturbation descriptors.
More recently, optimal transport has been introduced to model perturbation-induced shifts between unpaired cell populations. CellOT~\citep{bunne2023learning} learns a neural optimal-transport map from the control distribution to the perturbed distribution, whereas CellFlow~\citep{klein2025cellflow} combines OT-based cell pairing with
conditional flow matching. 
However, recent benchmarking studies have raised concerns about the robustness of such generalization. They show that on small-scale datasets, these methods tend to underperform simple baselines (\eg, mean-based predictors), particularly in unseen experimental settings~\citep{vinas2025systema, ahlmann2025deep, csendes2025benchmarking}.

In this study, we introduce \method, a flexible and scalable framework that captures single-cell perturbation effects across diverse, complex, and heterogeneous systems. It combines a conditional flow-based predictor with decoupled energy-based endpoint refinement, enabling low-cost and efficient extension to novel perturbation scenarios.
First, to address the unpaired nature of pre- and post-perturbation cellular profiles~\citep{schiebinger2019optimal, tong2020trajectorynet, bunne2023learning}, \method constructs cell-level paired data via optimal transport, which provides fine-grained supervision for model training. 
It then employs flow matching~\citep{lipman2023flow, tong2024improving} to model cellular perturbation responses as a continuous dynamical process, by learning velocity fields that characterize state transitions along pseudo-paired trajectories.
Building on this, our method incorporates multimodal condition inputs spanning cell type, drug identity and drug dosage 
to condition the flow model.
However, when generalized to perturbation conditions absent from the training data, such as unseen drug-dose conditions, this OT-guided conditional flow-matching backbone may produce biased initial predictions, a limitation also observed in existing methods.
To mitigate this bias, \method introduces energy-guided refinement by augmenting each Euler update with a correction derived from the gradient of a condition-specific energy function.
This strategy can be naturally extended to multiple conditions through a weighted sum of energy functions defined for individual independent conditions.

We evaluate \method across four distinct phenotypic and transcriptomic benchmarks: the imaging-based 4i dataset~\citep{bunne2023learning}, the transcriptomic SciPlex3 dataset~\citep{srivatsan2020massively}, the combinatorial ComboSciPlex dataset~\citep{lotfollahi2023cpa}, and the large-scale LINCS L1000 compendium~\citep{subramanian2017next}. 
Our evaluation follows a progressively challenging paradigm. 
We first validate the model's capacity to reconstruct observed perturbation effects. 
On 4i~\citep{bunne2023learning} and SciPlex3~\citep{srivatsan2020massively}, \method achieves low distributional discrepancies while preserving marker distributions, cell-state geometry, and condition-specific heterogeneity.
We then evaluate zero-shot generalization in out-of-distribution (OOD) settings. 
\method yields superior distribution-level fidelity on held-out drug-dose conditions, unseen cellular contexts, and novel drug combinations.
Next, we assess \method when only a limited number of cells from the target perturbation condition are available for training.
Given only three target-condition cells in 4i, \method substantially outperforms PRNet~\citep{qi2024prnet} and scDFM~\citep{yu2026scdfm} across all evaluated correlation and distributional metrics. 
Finally, we use L1000 to assess whether these statistical gains reflect biologically meaningful regulation~\citep{subramanian2017next}. 
We show that energy guidance improves directional concordance with L1000 consensus signatures for the most responsive genes, with gains observed in most evaluated contexts.
Across these scenarios, \method accurately predicts mean responses, models population heterogeneity, and reproduces biologically meaningful perturbation signatures in diverse experimental settings.

\begin{figure}[!th]
	\begin{center}
			\begin{overpic}[width=1\linewidth,percent]{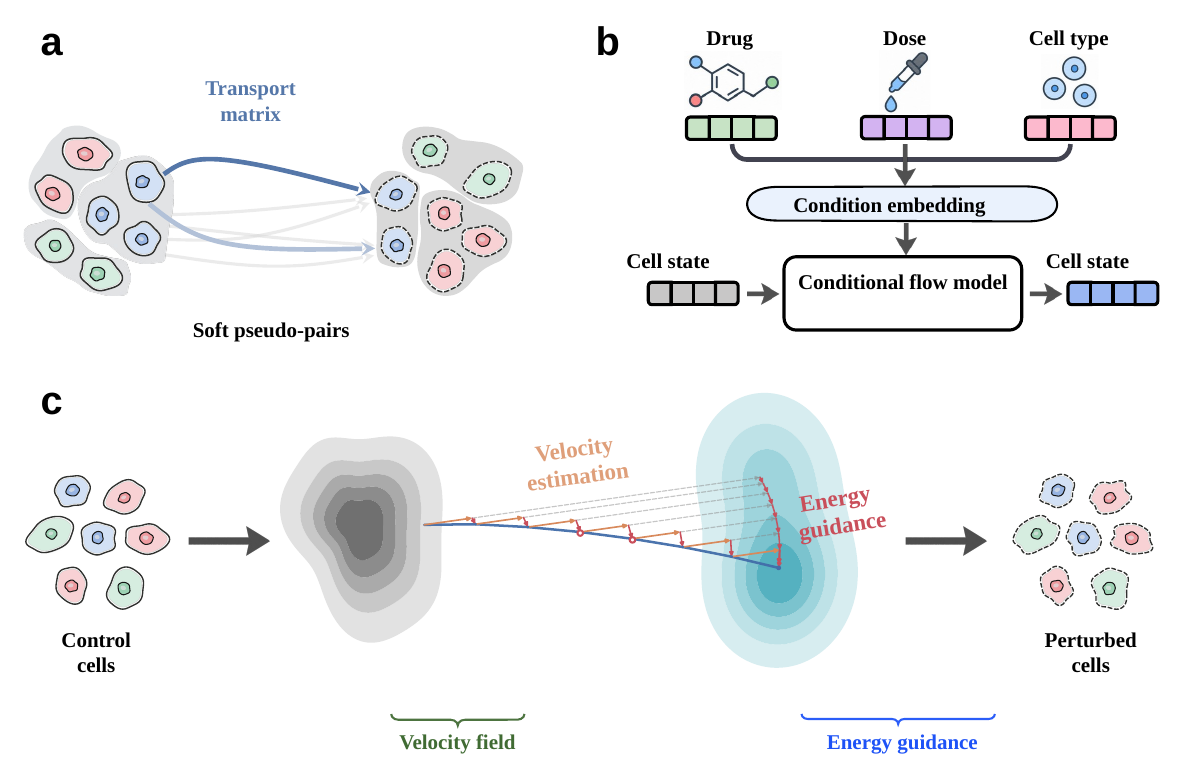}
				\put(11.4,51.5){\small $\bm{x}_i^{u}$}
				\put(25.8,54.5){\small \textcolor[rgb]{0.031,0.349,0.643}{$T^{*}$}}
				\put(29.2,50){\small $\bm{x}_j^{p}$}
				\put(28.8,40){\small $\bm{x}_{j+1}^{p}$}

				\put(83.1,45.8){\small $\bm{\xi}$}
				\put(60,41.2){\small $\bm{x}_t$}
				\put(95,41.2){\small $\bm{x}_{t+1}$}
				\put(71,37.2){\small $v_{\theta}(\bm{x}_t,t,\bm{\xi})$}

				\put(33,20){\small $\bm{x}^u$}
				\put(47.8,17.1){\small $\tilde{\bm{x}}_{t_k}$}
				\put(52.2,16.5){\small $\tilde{\bm{x}}_{t_{k+1}}$}
				\put(48.5,19.3){%
                    \rotatebox{20}{%
                        \scriptsize
                        \textcolor[rgb]{0.90,0.45,0.10}{$\Delta t v_{\theta}$}
                    }%
                }
                \put(52.8,19.2){%
                    \rotatebox{11}{%
                        \scriptsize
                        \textcolor[rgb]{0.788,0.310,0.357}{$\nabla E_{\phi}$}
                    }%
                }
				\put(65.2,14.8){\small $\bm{x}^{p}$}

				\put(16.1,5.2){\fontsize{9.5pt}{10pt}\selectfont $\hat{\bm{x}}_{t_{k+1}}=\tilde{\bm{x}}_{t_k}+\Delta t\,v_{\theta}(\tilde{\bm{x}}_{t_k},t_k,\bm{\xi})$}

				\put(49,5.2){\fontsize{9.5pt}{10pt}\selectfont $\tilde{\bm{x}}_{t_{k+1}}=\hat{\bm{x}}_{t_{k+1}}-\lambda_t\nabla_{\hat{\bm{x}}_{t_{k+1}}}E_{\phi}(\hat{\bm{x}}_{t_{k+1}},\bm{\xi})$}
		\end{overpic}
	\end{center}
	\vspace{-7pt}
	\small
    \caption{\textbf{Overview of \method.}
    \textbf{(a)} Optimal transport establishes soft correspondences between unpaired control and perturbed cells through the transport matrix $T^{*}$, yielding soft pseudo-pairs for training.
    \textbf{(b)} The perturbation vector $\bm{p}$ and biological context $\bm{c}$ are encoded as conditional inputs to the flow model, which predicts the velocity field $v_{\theta}(\bm{x}_{t},t,\bm{\xi})$ from the cell state $\bm{x}_{t}$ and time $t$.
    \textbf{(c)} During inference, cells are progressively transported toward the perturbed distribution by alternating flow-based velocity updates with energy-guided refinement. The former follows the learned conditional velocity field, whereas the latter adjusts cellular states along the energy gradient toward condition-consistent low-energy regions.}
    \label{fig:architecture}
	\vspace{-10pt}
\end{figure}

\section{Experiment}\label{exp}

\subsection{\method predicts perturbation responses with energy-guided flow matching}
\label{exp:method_overview}

Single-cell perturbation data contain independently sampled control and treated populations rather than matched pre- and post-perturbation cells. 
\method therefore formulates response prediction as conditional population transport: given control cells and a perturbation context, it generates a treated population that recovers both the mean response and cell-to-cell heterogeneity.
\method first constructs geometry-aware pseudo-pairs between control and treated cells using optimal transport. 
These pairs provide structured supervision for a conditional flow-matching model, which learns a continuous velocity field from baseline to perturbed states while conditioning on the perturbation and cellular context. 
The resulting flow captures the principal response modes and the global organization of the target population.
\method further calibrates the flow-matching trajectories through energy-guided refinement. 
At each Euler step, the gradient of a condition-specific energy function shifts the generated cell state toward response profiles that are more compatible with the queried perturbation.
This formulation extends naturally to perturbations specified by multiple attributes: guidance for drug identity, dosage, and other independent conditions can be combined through their respective energy terms.
This decouples population transport from endpoint calibration and permits condition-specific refinement without retraining the flow backbone.

We evaluate \method on four complementary benchmarks: the imaging-based 4i dataset~\citep{bunne2023learning}, the large-scale transcriptomic SciPlex3 dataset~\citep{srivatsan2020massively}, the combinatorial perturbation dataset ComboSciPlex~\citep{lotfollahi2023cpa}, and the LINCS L1000 perturbational transcriptomics resource~\citep{subramanian2017next}.
The evaluation first examines the reconstruction of observed perturbation responses on 4i and SciPlex3.
It then assesses generalization to previously unseen drug-dose conditions and cellular contexts in SciPlex3, as well as unseen drug combinations in ComboSciPlex.
We further evaluate response prediction on 4i when only a small number of cells from each target perturbation are available for training.
Finally, L1000 consensus signatures are used as an external reference to evaluate whether the predicted responses recover perturbation-induced directions of gene regulation.
Across these settings, \method remains competitive in mean-response prediction and shows consistently strong distributional recovery, while also preserving biologically meaningful transcriptional response directions.

\begin{figure}[!th]
	\begin{center}
		\includegraphics[width=1\linewidth]{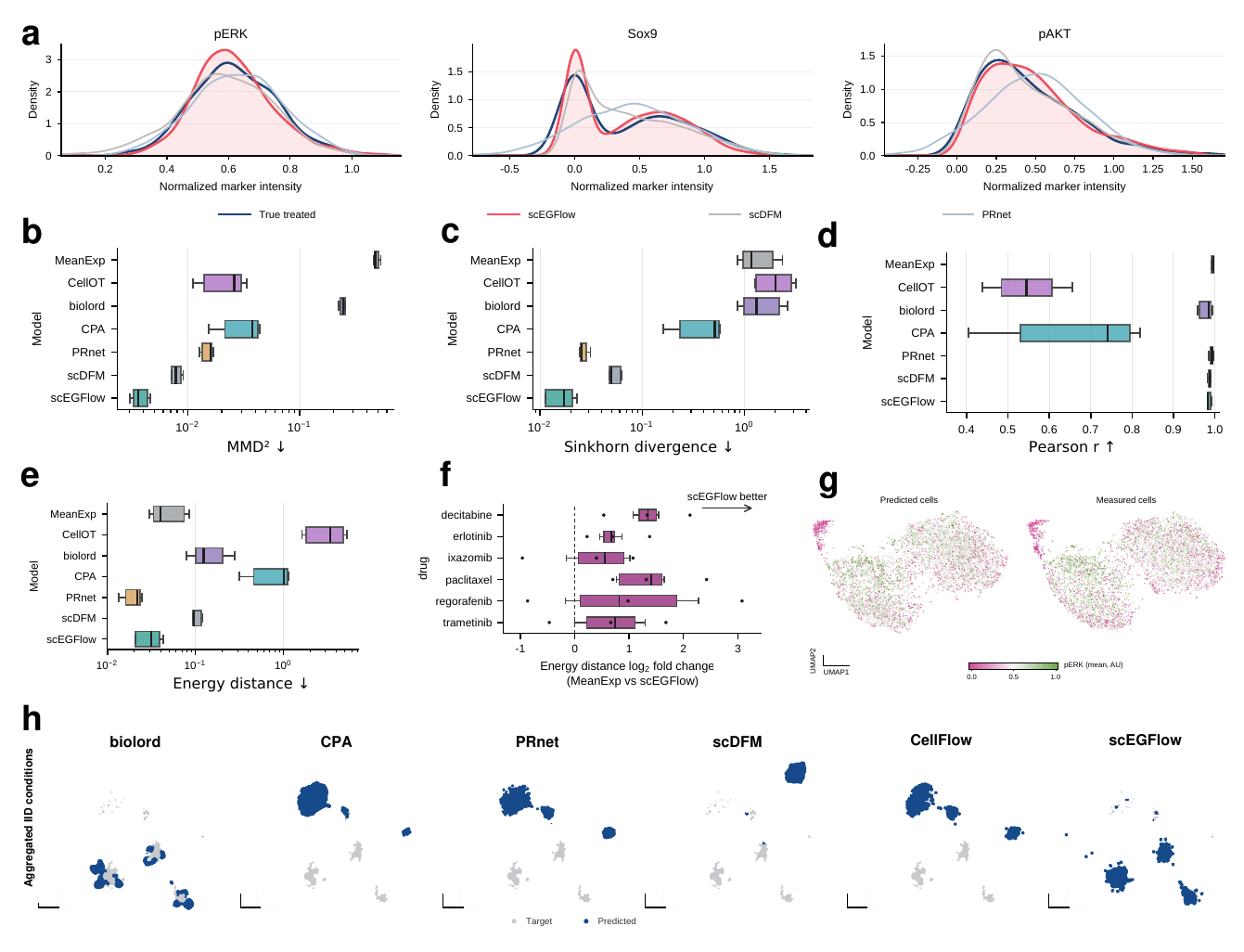}
	\end{center}
	\vspace{-7pt}
	\small
    \caption{\textbf{Reconstruction of single-cell perturbation responses across diverse data modalities.}
    \textbf{(a)} Marginal distributions of normalized pERK, Sox9, and pAKT marker intensities on the 4i~\citep{bunne2023learning} benchmark. Measured treated cells are shown in dark blue, and predictions from \method, scDFM, and PRNet are shown in red, gray, and light blue, respectively. 
    \textbf{(b--e)} IID reconstruction performance on the 4i benchmark, evaluated using MMD$^2$ \textbf{(b)}, Sinkhorn divergence \textbf{(c)}, Pearson correlation between condition-level mean profiles \textbf{(d)}, and energy distance \textbf{(e)}. Lower values are better for MMD$^2$, Sinkhorn divergence, and energy distance, whereas higher values are better for Pearson correlation.
    \textbf{(f)} Drug-level comparison between \method and the MeanExp baseline for six selected 4i perturbations. The horizontal axis shows the log$_2$ ratio of the energy distances obtained by MeanExp and \method.
    \textbf{(g)} UMAP visualization of predicted and measured 4i cells aggregated across all evaluated drug treatments in a shared phenotypic state space. Cells are colored by normalized mean pERK intensity. AU, arbitrary units.
    \textbf{(h)} UMAP comparison of SciPlex3~\citep{srivatsan2020massively} IID response reconstruction across aggregated drug-dose conditions. Measured target cells are shown in gray and predicted cells from each method are shown in blue.}
    \label{fig:reconstruction}
	\vspace{-10pt}
\end{figure}

\subsection{Predicting single-cell perturbation under various and complex scenarios}
\label{sec:iid_prediction}

In single-cell perturbation experiments, cellular responses vary with treatment, cell type, baseline cellular state, and other factors, resulting in diverse and complex response patterns. 
To evaluate whether our method can accurately predict post-perturbation cell populations under such settings, we conduct experiments on two single-cell perturbation benchmarks, 4i~\citep{bunne2023learning} and SciPlex3~\citep{srivatsan2020massively}, which differ in readout modality and condition space.

We first evaluate our method on the 4i benchmark, which contains imaging-based phenotypic profiles from a co-culture of two patient-derived melanoma cell lines, M130219 and M130429, under diverse drug perturbations. 
Fig.~\ref{fig:reconstruction}a compares the measured and predicted marginal distributions of three representative markers with distinct distributional characteristics~\citep{bunne2023learning}.
Although all methods capture the overall response trends of these features, scDFM and PRNet exhibit noticeable distributional shifts or over-smoothing, while \method more closely follows the measured distributions, preserving heterogeneous and accurate population responses.
Fig.~\ref{fig:reconstruction}b-e compare \method with other methods in terms of MMD$^2$, Sinkhorn divergence, Pearson correlation, and energy distance. 
Together with Table~\ref{tab:iid_4i_sciplex3_results}, these results show that \method achieves the lowest MMD$^2$ and Sinkhorn divergence on the 4i~\citep{bunne2023learning} benchmark, while maintaining competitive energy distance and Pearson correlation.
We further test whether \method captures drug-specific responses.  
Motivated by recent findings that simple mean-based baselines can be difficult to outperform in single-cell perturbation prediction~\citep{ahlmann2025deep}, we compare \method with MeanExp across six drug perturbations from the 4i~\citep{bunne2023learning} dataset. 
As shown in Fig.~\ref{fig:reconstruction}f, \method achieves a consistent improvement in energy distance for all six drugs,
demonstrating that it better preserves drug-dependent population variation.
Fig.~\ref{fig:reconstruction}g visualizes the reconstructed 4i populations by embedding predicted and measured cells in a shared UMAP space and coloring them by pERK intensity.
The cell distributions predicted by our method show qualitative overlap with the measured population in the shared UMAP embedding, together with a similar spatial organization of low- and high-pERK states across the embedding. 
This agreement indicates that \method preserves both the global arrangement of phenotypic states and the signaling-associated variation within the treated population.

We then extend the evaluation to SciPlex3~\citep{srivatsan2020massively}, a substantially larger transcriptomic benchmark spanning multiple cancer cell lines, compounds, and dosage. 
As shown in Table~\ref{tab:iid_4i_sciplex3_results}, \method achieves the lowest MMD$^2$, energy distance and Sinkhorn divergence while maintaining high Pearson correlation. 
Thus, the reconstruction advantage observed in the imaging-based 4i~\citep{bunne2023learning} setting remains evident in a higher-dimensional gene-expression space with more diverse cellular and perturbation contexts.
Figure~\ref{fig:reconstruction}h provides a complementary view of the SciPlex3 results by jointly embedding predicted and measured cells across aggregated conditions. 
Several baselines concentrate their predictions in a limited subset of the target-supported regions or form visibly separated clusters, suggesting incomplete coverage of the measured response distribution. 
In contrast, \method predictions are distributed across multiple target regions and show broader overlap with the measured cells. 
This pattern is consistent with the quantitative results, indicating that \method more effectively recovers the diversity of transcriptomic response states rather than reproducing only a small number of dominant modes.

Together, these results show that \method reconstructs perturbation responses beyond condition-level mean matching by preserving condition-specific population structure, heterogeneous response modes, and the organization of
cell states. 
This consistent behavior highlights the ability of \method to recover both average treatment effects
and within-condition response heterogeneity.

\begin{table*}[t]
\centering
\caption{IID perturbation prediction on 4i~\citep{bunne2023learning}
and SciPlex3~\citep{srivatsan2020massively}.}
\label{tab:iid_4i_sciplex3_results}

\begin{threeparttable}
\small
\setlength{\tabcolsep}{5pt}
\renewcommand{\arraystretch}{1.08}

\begin{tabular}{@{}l|cccc|cccc@{}}
\toprule
\multirow{2}{*}{Method}
& \multicolumn{4}{c|}{4i}
& \multicolumn{4}{c}{SciPlex3} \\

& Energy$\downarrow$
& Sinkhorn$\downarrow$
& MMD$^2$$\downarrow$
& Pearson$\uparrow$
& Energy$\downarrow$
& Sinkhorn$\downarrow$
& MMD$^2$$\downarrow$
& Pearson$\uparrow$ \\
\midrule

MeanExp
& 0.0510
& 1.4463
& 0.4839
& \textbf{0.9925}
& 2.2626
& 12.7494
& 0.6887
& 0.9914 \\

CellOT~\citep{bunne2023learning}
& 3.1399
& 2.0230
& 0.0199
& 0.5513
& --
& --
& --
& -- \\

biolord~\citep{piran2024disentanglement}
& 0.1507
& 1.2904
& 0.2411
& 0.9669
& 1.1494
& 0.9318
& 0.2192
& 0.9949 \\

CPA~\citep{lotfollahi2023cpa}
& 0.6125
& 0.3091
& 0.0257
& 0.7455
& 2.9388
& 1.5674
& 0.2509
& 0.9476 \\

PRNet~\citep{qi2024prnet}
& \textbf{0.0208}
& 0.0254
& 0.0145
& 0.9910
& 0.6419
& 0.3788
& 0.0145
& 0.9856 \\

scDFM~\citep{yu2026scdfm}
& 0.1035
& 0.0541
& 0.0078
& 0.9831
& 21.9521
& 11.6483
& 0.0979
& 0.8565 \\

CellFlow~\citep{klein2025cellflow}
& --
& --
& --
& --
& 1.0165
& 0.8560
& 0.0171
& \textbf{0.9955} \\

\rowcolor{gray!12}
\method
& 0.0303
& \textbf{0.0162}
& \textbf{0.0035}
& 0.9868
& \textbf{0.5215}
& \textbf{0.3100}
& \textbf{0.0140}
& 0.9887 \\

\bottomrule
\end{tabular}
\end{threeparttable}
\end{table*}

\begin{table*}[t]
\centering
\caption{OOD perturbation prediction under unseen drug-dose conditions
and in an unseen cellular context on SciPlex3~\citep{srivatsan2020massively}.}
\label{tab:ood_sciplex3_results}

\begin{threeparttable}
\small
\setlength{\tabcolsep}{5pt}
\renewcommand{\arraystretch}{1.08}

\begin{tabular}{@{}l|cccc|cccc@{}}
\toprule
\multirow{2}{*}{Method}
& \multicolumn{4}{c|}{OOD drug-dose}
& \multicolumn{4}{c}{OOD cellular context} \\

& Energy$\downarrow$
& Sinkhorn$\downarrow$
& MMD$^2$$\downarrow$
& Pearson$\uparrow$
& Energy$\downarrow$
& Sinkhorn$\downarrow$
& MMD$^2$$\downarrow$
& Pearson$\uparrow$ \\
\midrule

MeanExp
& 0.8735
& 5.7422
& 0.6824
& 0.9966
& 52.9832
& 27.5550
& 0.7647
& 0.8090 \\

biolord~\citep{piran2024disentanglement}
& 0.7792
& 0.5939
& 0.2114
& 0.9966
& 51.4051
& 26.7684
& 0.7584
& 0.8154 \\

CPA~\citep{lotfollahi2023cpa}
& 2.2859
& 1.2211
& 0.2385
& 0.9607
& 14.9031
& 7.5212
& 0.4386
& 0.6901 \\

PRNet~\citep{qi2024prnet}
& 0.4077
& 0.2211
& 0.0051
& 0.9909
& 55.4931
& 22.1669
& 0.1777
& 0.8003 \\

CellFlow~\citep{klein2025cellflow}
& 0.5218
& 0.3692
& 0.0065
& \textbf{0.9977}
& 22.3046
& 12.2336
& 0.1624
& \textbf{0.9254} \\

scDFM~\citep{yu2026scdfm}
& 68.4362
& 34.5800
& 0.1501
& 0.8674
& 54.7546
& 32.9439
& 0.2249
& 0.6487 \\

\rowcolor{gray!12}
\method
& \textbf{0.2438}
& \textbf{0.1391}
& \textbf{0.0039}
& 0.9945
& \textbf{3.8726}
& \textbf{2.0344}
& \textbf{0.0831}
& 0.9253 \\

\bottomrule
\end{tabular}
\end{threeparttable}
\end{table*}



\begin{figure}[!th]
	\begin{center}
		\includegraphics[width=0.8\linewidth]{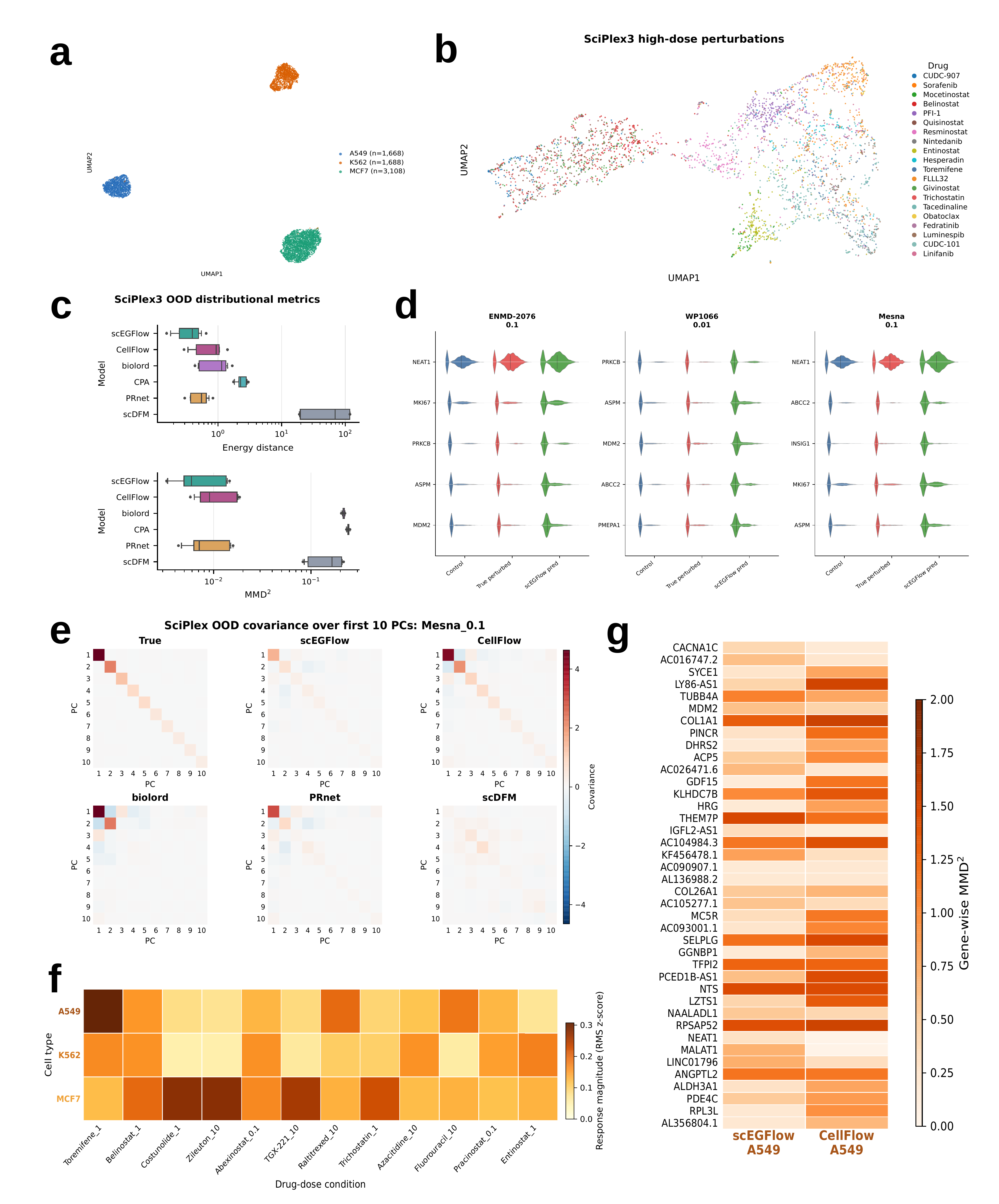}
	\end{center}
	\vspace{-15pt}
	\small
    \caption{\textbf{Generalization of single-cell perturbation responses across unseen chemical and cellular contexts.}
    \textbf{(a)} UMAP visualization of unperturbed cells colored by three cell types (A549, K562, and MCF7 cells).
    \textbf{(b)} UMAP visualization of measured MCF7 cells treated with 20 representative compounds at the highest profiled dose.
    \textbf{(c)} Distribution-level prediction performance for held-out SciPlex3~\citep{srivatsan2020massively} drug-dose conditions. Boxplots compare the energy distance and MMD$^2$ between predicted and measured target-cell populations over the OOD test set.
    \textbf{(d)} Gene-expression distributions of the five most strongly perturbed genes for three held-out drug-dose conditions, comparing control, measured perturbed, and \method-predicted cells.
    \textbf{(e)} Covariance matrices over the first 10 principal components for the held-out Mesna\_0.1 condition, comparing the measured target population with populations predicted by \method and the baseline methods; all six matrices are computed in the same PCA basis fitted on the training cells.
    \textbf{(f)} Heatmap of the measured treated-versus-control response magnitude
    across representative drug-dose conditions in three cell types.
    Response magnitude is computed as the root-mean-square of the
    gene-wise treated-minus-control mean expression shifts after z-score
    standardization over the retained genes.
    \textbf{(g)} Gene-wise MMD$^2$ for \method and CellFlow on the most strongly perturbed genes, selected by the absolute treated-minus-control mean shift in the held-out A549 context. For each gene, cells across all evaluated drug-dose conditions are pooled before computing MMD$^2$ between measured and predicted expression distributions. Lower values indicate better agreement.
    }
    \label{fig:OOD}
	\vspace{-10pt}
\end{figure}

\subsection{Generalizing cellular responses across unseen chemical and cellular contexts}\label{sec:ood_prediction}

We next assess whether \method can extrapolate from observed patterns to previously unseen combinations of experimental conditions, \ie, out-of-distribution (OOD) settings.
This capability is evaluated on SciPlex3~\citep{srivatsan2020massively} and ComboSciPlex~\citep{lotfollahi2023cpa}. 
SciPlex3 exhibits substantial heterogeneity across cell types and single-drug dose conditions (Fig.~\ref{fig:OOD}a,b), whereas ComboSciPlex~\citep{lotfollahi2023cpa} includes both monotherapy and combination-drug treatments.
Specifically, we consider three held-out settings: unseen drug-dose conditions, unseen cellular contexts, and unseen drug combinations.
Together, these settings evaluate \method's generalization across chemical, cellular, and combinatorial contexts.

\textbf{Unseen drug-dose conditions.}
We construct the test data by withholding all perturbed cells from three drug-dose conditions (\ie, ENMD-2076 at 0.1~\(\mu\)M, WP1066 at 0.01~\(\mu\)M, and Mesna at 0.1~\(\mu\)M) from training on SciPlex3~\citep{srivatsan2020massively}. 
\method is used to predict responses to these unseen conditions, which are evaluated against the corresponding measured cell populations.
As shown in Fig.~\ref{fig:OOD}c and Table~\ref{tab:ood_sciplex3_results}, \method achieves the lowest energy distance, Sinkhorn divergence, and MMD$^2$ in this split, while maintaining a high Pearson correlation, indicating superior population-level agreement with the measured responses.
This result is further supported by the gene-level distributions in Fig.~\ref{fig:OOD}d, where \method faithfully captures both the expression shifts relative to the control and the specific distributional shapes of the five most strongly perturbed genes across the three held-out conditions.
Fig.~\ref{fig:OOD}e provides a detailed analysis of the held-out Mesna at 0.1~\(\mu\)M condition by comparing the covariance structures of the measured and predicted populations over the first 10 principal components.
\method successfully captures the dominant covariance patterns and preserves the primary structure of response heterogeneity. 
Although it does not achieve the absolute best match on this specific second-order statistic, it remains highly competitive.


\textbf{Unseen cellular contexts.}
Cellular context strongly shapes transcriptional responses to chemical perturbations~\cite{mcfarland2020multiplexed}, as evidenced by the markedly different response magnitudes observed among A549, K562, and MCF7 cells under the same
drug-dose conditions in SciPlex3~\citep{srivatsan2020massively} (Fig.~\ref{fig:OOD}f).
We therefore next evaluate whether \method can extrapolate learned perturbation patterns to a novel cellular context by withholding all perturbed A549 profiles during training on SciPlex3.
As detailed in Table~\ref{tab:ood_sciplex3_results}, \method sustains the performance advantage observed in the unseen cellular-context evaluation.
It achieves the strongest overall agreement with the measured A549 response populations on distribution-level metrics, while maintaining a competitive mean-profile correlation.
The gene-level analysis in Fig.~\ref{fig:OOD}g further compares \method and CellFlow on the most strongly perturbed genes.
\method yields lower MMD$^2$ for the majority of these genes, indicating that its advantage extends beyond aggregate population scores to the accurate reconstruction of perturbation-responsive gene-expression distributions.

\begin{table}[t] 
\centering 

\caption{OOD perturbation prediction on ComboSciPlex~\citep{lotfollahi2023cpa}.}
\label{tab:ood_combosciplex_result} 
\begin{threeparttable} 
\setlength{\tabcolsep}{8pt} 
\renewcommand{\arraystretch}{1.12} 
\begin{tabular}{l|cccc} 
\toprule Method & Energy$\downarrow$ & Sinkhorn$\downarrow$ & MMD$^2$$\downarrow$ & Pearson$\uparrow$ \\
\midrule CellOT~\citep{bunne2023learning} & 28.19 & 112.71 & 0.0262 & 0.9625 \\
biolord~\citep{piran2024disentanglement} & 37.98 & 98.97 & 0.6087 & 0.9522 \\
CPA~\citep{lotfollahi2023cpa} & 29.09 & 94.05 & 0.3239 & 0.9608 \\
PRNet~\citep{qi2024prnet} & 50.59 & 202.66 & \textbf{0.0259} & 0.9337 \\
CellFlow~\citep{klein2025cellflow} & 25.06 & 93.52 & 0.0664 & \textbf{0.9692} \\
scDFM~\citep{yu2026scdfm} & 91.23 & 246.80 & 0.0354 & 0.8848 \\
\rowcolor{gray!12}
\method& \textbf{24.86} & \textbf{93.02} & 0.0527 & 0.9667 \\
\bottomrule 
\end{tabular} 
\end{threeparttable} 
\end{table}

\begin{figure}[!t]
	\begin{center}
		\includegraphics[width=1\linewidth]{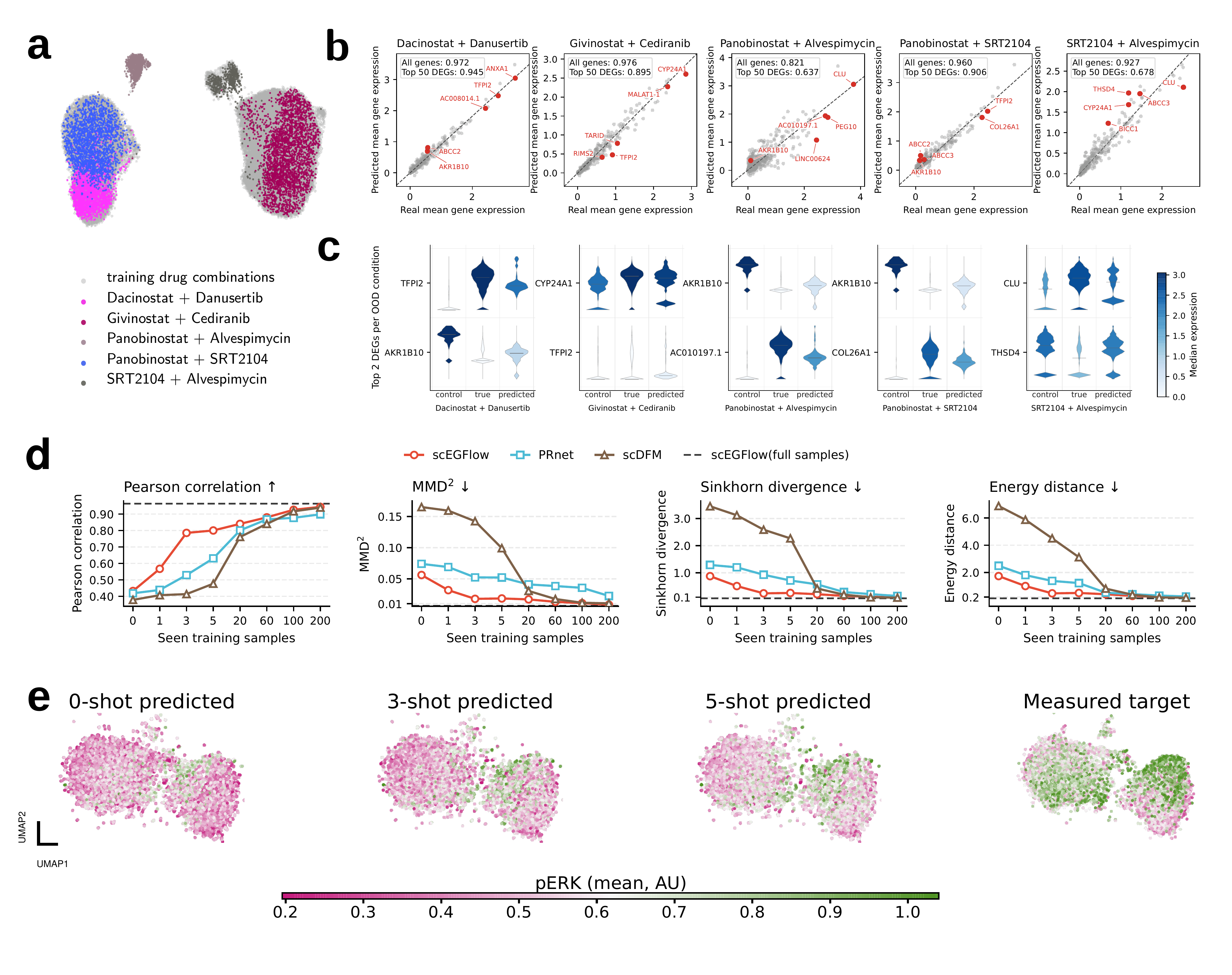}
	\end{center}
	\vspace{-7pt}
	\small
	\captionsetup{font=small}
	\caption{\textbf{Generalization to unseen drug combinations and sample-efficient prediction.}
	\textbf{(a)} UMAP visualization of measured ComboSciPlex~\citep{lotfollahi2023cpa} cells, showing the five held-out drug-combination responses and their positions relative to the response manifold.
	\textbf{(b)} Predicted versus measured mean gene expression for the five held-out drug combinations, shown for all genes and the top 50 differentially expressed genes (DEGs).
	\textbf{(c)} Expression distributions of the top two DEGs for each held-out drug combination, comparing control, measured treated, and \method-predicted cells.
	\textbf{(d)} Few-shot prediction on the 4i~\citep{bunne2023learning} benchmark under different numbers of target-condition training cells, evaluated using Pearson correlation, MMD$^2$, Sinkhorn divergence, and energy distance.
	\textbf{(e)} UMAP comparison of zero-shot predictions, 3-shot predictions, 5-shot predictions and measured target cells, colored by normalized pERK intensity.}
	\vspace{-16pt}
	\label{fig:few_shot_combo}
\end{figure}

\textbf{Unseen combinatorial perturbations.}
Drug combinations are widely used to improve the likelihood and
magnitude of therapeutic responses and to reduce acquired
resistance~\citep{palmer2017combination}.
However, combined treatments can produce non-additive transcriptional responses that cannot be reliably inferred from either drug alone \citep{lotfollahi2023cpa}. 
Finally, we evaluate whether \method can predict the single-cell response distributions of unseen drug combinations on the ComboSciPlex benchmark~\citep{lotfollahi2023cpa}.
We use an OOD split in which all cells from five target combinations are excluded from training: Dacinostat + Danusertib, Givinostat + Cediranib, Panobinostat + Alvespimycin, Panobinostat + SRT2104, and SRT2104 + Alvespimycin.
This capability relies on the assumption that responses to unseen drug combinations can be approximated from combinations observed during training.
As illustrated in Fig.~\ref{fig:few_shot_combo}a, the measured response populations of these combinations exhibit varying degrees of overlap with the training conditions, which reflects their divergence from the training distribution and the difficulty of extrapolating to novel combinations.
Table~\ref{tab:ood_combosciplex_result} summarizes prediction performance on the pooled OOD test set comprising the five held-out combinations.
scEGFlow achieves the lowest energy distance and Sinkhorn divergence while maintaining competitive Pearson correlation, although its MMD$^2$ is higher than those of PRNet and CellOT.
Fig.~\ref{fig:few_shot_combo}b complements this aggregate analysis by comparing predicted and measured mean-expression profiles for each combination.
\method shows strong agreement across most combinations for both all genes and the top 50 differentially expressed genes.
Fig.~\ref{fig:few_shot_combo}c further examines the two most strongly affected genes in each condition at single-cell resolution. 
The predicted distributions generally reproduce the direction and magnitude of the measured shifts from control to perturbed cells, indicating that \method learns the condition-specific perturbation trends rather than merely matching the overall expression level.

Collectively, these OOD evaluations demonstrate that \method generalizes to unseen drug-dose conditions, cellular contexts, and drug combinations.
This capability can help prioritize experimental validation across diverse perturbations and cell lines, thereby reducing the burden of large-scale drug screening.

\subsection{Adapting to target conditions with limited observations}

\label{sec:few_shot_prediction}

Although generalization to entirely unseen perturbations is appealing, existing methods, including \method, still fall short of fully supervised performance. 
A practical and efficient compromise is to adapt the model using a small number of target-condition observations. 
To evaluate this setting on the 4i~\citep{bunne2023learning} benchmark, we select five target drugs (\ie, cisplatin, crizotinib,
dabrafenib, dacarbazine, and dasatinib). 
For each target condition, we allocate 1, 3, 5, 20, 60, 100 or 200 cells for training, while reserving the remaining cells for evaluation.

%
As illustrated in Fig.~\ref{fig:few_shot_combo}d, \method consistently achieves the best performance across different levels of target-condition supervision, with this advantage being particularly pronounced in highly data-constrained regimes.
Remarkably, even a single target-condition cell provides a clear performance gain over the zero-shot setting, indicating that \method can extract valuable condition-specific information from minimal supervision.
The benefit becomes substantially more evident when three target cells are available.
Specifically, given only three samples, our model markedly outperforms PRNet~\citep{qi2024prnet} and scDFM~\citep{yu2026scdfm} across all evaluated metrics, including Pearson correlation (0.7857 vs. 0.5293 and 0.4137), MMD$^2$ (0.0179 vs. 0.0523 and 0.1425), Sinkhorn divergence (0.2402 vs. 0.9262 and 2.5888), and energy distance (0.4735 vs. 1.3850 and 4.5085). 
As more target-condition observations become available, the performance gap between methods narrows, yet \method maintains its overall superiority.
Such robustness in the few-sample regime demonstrates that \method can effectively exploit scarce observations to capture both mean perturbation effects and population-level distributions.

Fig.~\ref{fig:few_shot_combo}e further reveals what is gained from limited target-condition supervision. 
In the zero-shot setting, \method captures the overall cell-state manifold, but the distribution of pERK activity is only partially aligned with that of the measured target population.
With only 3 target-condition cells, the predicted population more closely reproduces the organization of low- and high-pERK regions across the manifold.
This improvement suggests that providing \method with even a small number of target-specific observations enables it to calibrate condition-dependent signaling patterns and better recover response heterogeneity.



\subsection{Recapitulating directional concordance with L1000 perturbational signatures}

In perturbational transcriptomics, cellular responses are conventionally characterized by the direction of gene-expression changes, where up- and down-regulated genes together form a signed signature~\citep{lamb2006connectivity,subramanian2017next,peng2026generative}. 
Building on this concept, we investigate whether \method can accurately recover the directionality of these induced responses.

To this end, we evaluate \method on LINCS L1000~\citep{subramanian2017next}, a large-scale transcriptomics dataset profiling cellular responses to chemical and genetic perturbations. 
L1000 measurements are processed through five data levels, ranging from raw measurements to consensus signatures. 
Here, we utilize Levels 3 and 5 to construct our training and evaluation setup.
Level 3 provides normalized gene-expression profiles for individual control and compound-treated samples, whereas Level 5 aggregates replicate profiles to represent the consensus perturbation signatures induced by each perturbation.
Consequently, we train \method on the Level 3 expression profiles and compare its predicted response directions against the corresponding Level 5 signatures. 

\label{sec:l1000_biological_response}
Given the scale of the full L1000 compendium~\citep{subramanian2017next}, we construct a focused subset of compound perturbations by pairing Level 3 profiles with their corresponding Level 5 signatures. 
The resulting dataset contains $24{,}104$ Level 3 expression profiles over $978$ landmark genes, spanning $21$ cell lines and $79$ compounds. 
We hold out six drug–dose conditions for OOD evaluation, ultimately yielding $111$ evaluation contexts after pairing with the corresponding Level 5 signatures. 

For each context, we rank the $978$ landmark genes by the magnitude of their Level 5 expression changes, retaining the top $10\%$ to yield $98$ genes with the strongest perturbation effects. 
We quantify \emph{directional response concordance} as the fraction of these top genes for which the predicted outcome and the corresponding Level 5 signature agree on the direction of regulation.

\begin{figure}[!t]
	\begin{center}
		\includegraphics[width=1\linewidth]{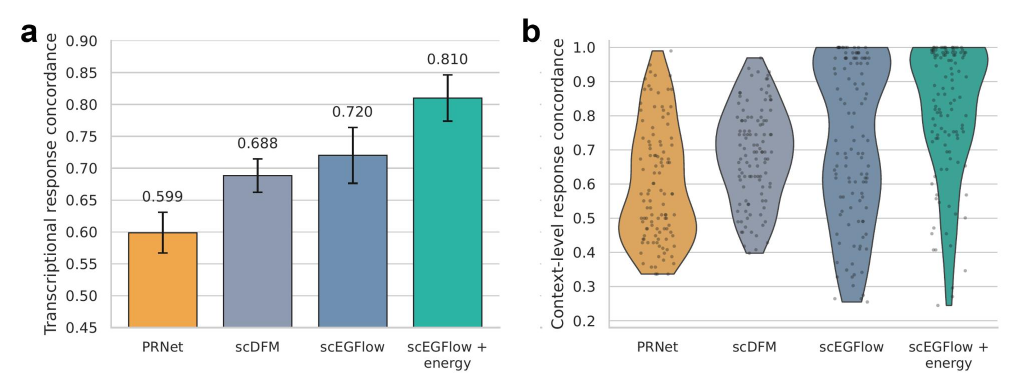}
	\end{center}
	\vspace{-7pt}
	\small
	\captionsetup{font=small}
    \caption{\textbf{Directional concordance with L1000~\citep{subramanian2017next} perturbational signatures.}
    \textbf{(a)} Mean directional concordance between predicted perturbation responses and the corresponding L1000 Level 5 consensus signatures.
    \textbf{(b)} Context-level distribution of directional concordance across all $111$ evaluation contexts. Each point represents one specific cell line-drug-dose condition.}
	\label{fig:directional_concordance}
\end{figure}

As shown in Fig.~\ref{fig:directional_concordance}a, \method achieves higher directional concordance than PRNet~\citep{qi2024prnet} and scDFM~\citep{yu2026scdfm} even without energy guidance, demonstrating that the flow backbone inherently captures transcriptional shifts more faithfully. Incorporating energy-guided refinement further boosts the mean concordance from $72.0\%$ to $81.0\%$. 
This indicates that the learned energy function provides condition-specific information associated with improved directional agreement, enabling a more accurate reconstruction of the up- and down-regulation patterns encoded by the L1000 consensus signatures.

Fig.~\ref{fig:directional_concordance}b further details the concordance distribution across the $111$ evaluation contexts. 
Notably, energy guidance enhances predictive accuracy in $98$ of these contexts, confirming that the performance gain is not an artifact of a few outliers, but a robust phenomenon observed across diverse cell lines, compounds, and doses. 
Together, these results support \method's ability to capture the direction of gene-expression changes in the evaluated settings and show that energy guidance improves concordance with L1000 consensus signatures.

\section{Conclusion and discussion}

In this work, we present \method to predict single-cell perturbation responses as conditional transport between unpaired control and treated populations. 
\method combines optimal-transport-based soft coupling with conditional flow matching to learn population transitions, and leverages condition-specific energy gradients to refine the generated trajectories. 
This decoupling of transport learning from condition-specific calibration is the core design principle of the framework: the flow model captures the global dynamics of the response distribution, whereas energy guidance steers predictions toward states that are most compatible with the queried perturbation context. 
Moreover, energy terms representing independent attributes can be compositionally combined, providing a modular mechanism to incorporate drug, dose, and cellular context into the refinement process.

Evaluation across 4i~\citep{bunne2023learning}, SciPlex3~\citep{srivatsan2020massively}, ComboSciPlex~\citep{lotfollahi2023cpa}, and LINCS L1000~\citep{subramanian2017next} supports three main conclusions. 
First, \method accurately reconstructs observed perturbation responses across phenotypic and transcriptomic modalities, capturing both mean effects and within-condition heterogeneity.
Second, its distributional performance remains favorable in the evaluated held-out scenarios, encompassing held-out drug–dose conditions, an unseen cellular context, and novel drug combinations. 
In data-scarce regimes, \method readily leverages limited target observations to refine its predictions.
Third, driven by energy guidance, \method faithfully reproduces the up- and down-regulation patterns recorded in L1000 consensus signatures, demonstrating that its effectiveness stems from genuine biological concordance.
Together, these results show that \method successfully bridges statistical distributional alignment with biological fidelity in single-cell response modeling.

Several limitations define the scope of the present study. 
First, our method relies on an iterative optimization procedure, in which each step requires velocity prediction from the flow model and a derivative operation for the energy function.
As a result, our method incurs a higher computational cost than one-shot approaches. 
Second, our flow model adopts a simple linear interpolation path between each source cell and its perturbed counterpart, which simplifies biological transitions that may otherwise be nonlinear; furthermore, the OT-derived pseudo-pairs depend on feature-space distances and may not always reflect true cellular trajectories.
Third, energy guidance with multiple conditions assumes that all conditions are independent; when correlations exist between conditions, this assumption may lead to suboptimal results.

Looking ahead, several important directions could further advance \method for single-cell perturbation response prediction.
The modular design of \method offers the potential to reduce the computational complexity of extending prediction to additional perturbation conditions, which could open the door to the scalable exploration of larger and more diverse biological datasets.

Building upon this efficiency, future work could focus on extending the framework to more diverse perturbation types, cellular contexts, and experimental platforms.
Furthermore, future work could explore whether additional energy functions can extend \method to other perturbation types, such as gene knockouts~\citep{dixit2016perturbseq} and cytokine stimulation~\citep{kang2018multiplexed}, without retraining the flow backbone.
Thus, another promising direction is to improve the design of the energy function, for example by using off-the-shelf pretrained models to encode conditions and provide stronger priors.

\section{Methods}\label{sec11}

\subsection{Problem setup}\label{method:problem_setup}
To streamline model training and optimization, we predefine a set of cellular attributes, denoted as $\mathcal{G} = \{g_1, g_2, \dots, g_G\}$, consisting of profiled genes as well as cellular and nuclear morphological measurements.
The state of a cell $x$ is characterized by its profile across this set, represented as a vector $\bm{x} = [c_x(g_1), \dots, c_x(g_D)]^T \in \mathbb{R}^{D}$, where $c_x(g_i)$ signifies the quantified value of the $i$-th attribute. 
Our task is to predict the perturbed cell state $\bm{x}^p$ from an unperturbed baseline state $\bm{x}^u$, conditioned on a specific perturbation and a biological context.
Here, the perturbation is represented by a vector $\bm{p} \in \mathbb{R}^{M_d}$, where $M_d$ is the total number of unique perturbations in the vocabulary; the element $p_{m}$ specifies the dosage of the $m$-th perturbation, falling to zero ($p_{m}=0$) in its absence. 
The biological context is represented by a context embedding $\bm{c}$ derived from the corresponding unperturbed source population. In our implementation, $\bm{c}$ is defined as the mean control-cell state for each cell line.
We denote the complete condition by $\bm{\xi}=(\bm{p},\bm{c})$.
Formally, we aim to learn a conditional distributional predictor $f_\theta(\bm{x}^p \mid \bm{x}^u, \bm{\xi})$ that captures the distribution of perturbed cell states conditioned on baseline expression, perturbation, and biological context. 

\subsection{Optimal transport for pairing cells}\label{method:ot}

As measuring cellular perturbation responses requires destructive sampling,
the same cell cannot be assayed twice, resulting in unpaired distributions of
perturbed and unperturbed cells
\citep{schiebinger2019optimal, bunne2023learning}.
However, our proposed flow matching strategy requires paired data to learn the
conditional vector fields that connect the two distributions.
To address this, we employ optimal transport (OT)
\citep{cuturi2013sinkhorn, tong2024improving}
to establish soft correspondences between unperturbed and perturbed cells.

Formally, we denote the source and target distributions of control and perturbed
cells by $\mu$ and $\nu$, respectively, both defined on $\mathbb{R}^{D}$.
In each batch, we sample unperturbed cells
$\{\bm{x}_i^u\}_{i=1}^{N}\sim\mu$
and perturbed cells
$\{\bm{x}_j^p\}_{j=1}^{M}\sim\nu$.
The pairwise cost matrix $\bm{C}\in\mathbb{R}^{N\times M}$ is defined as
\begin{align}
C_{ij}
=
\|\bm{x}_i^u-\bm{x}_j^p\|_2^2.
\end{align}

We assign equal probability mass to all sampled cells and define the admissible
couplings as
\begin{align}
\Pi_{N,M}
=
\{
\bm{T}\in\mathbb{R}_{+}^{N\times M}:
\bm{T}\bm{1}_M
=
\frac{1}{N}\bm{1}_N,
\quad
\bm{T}^{\top}\bm{1}_N
=
\frac{1}{M}\bm{1}_M
\}.
\label{eq:ot_sub}
\end{align}
The optimal coupling $\bm{T}^{\star}$ is obtained by solving the entropically
regularized OT problem
\begin{align}
\bm{T}^{\star}
=
\operatorname*{arg\,min}_{\bm{T}\in\Pi_{N,M}}
\langle \bm{T},\bm{C}\rangle
+
\varepsilon
\sum_{i,j}
T_{ij}(\log T_{ij}-1),
\end{align}
where $\varepsilon$ controls the strength of entropic regularization and
$T_{ij}^{\star}$ denotes the coupling weight between
$\bm{x}_i^u$ and $\bm{x}_j^p$.
In practice, the mini-batch OT coupling is approximated using the Sinkhorn
algorithm with 100 iterations, with dataset-specific regularization strengths
reported in the experimental setup.

Based on $\bm{T}^{\star}$, we construct pseudo-paired samples for flow matching.
A straightforward strategy is to pair each source cell with the target cell
having the largest coupling weight.
However, such hard assignment collapses the soft coupling into a deterministic
match and ignores the remaining mass distribution
\citep{kim2023swamp}.
We therefore sample target cells according to the OT coupling.
For each source cell $\bm{x}_i^u$, the $i$-th row of $\bm{T}^{\star}$ is
normalized as
\begin{align}
q_{ij}
=
\frac{T_{ij}^{\star}}
{\sum_{\ell=1}^{M}T_{i\ell}^{\star}}.
\end{align}
We then sample
$j_i\sim\mathrm{Categorical}(q_{i1},\ldots,q_{iM})$
and construct the pseudo-pair
$(\bm{x}_i^u,\bm{x}_{j_i}^p)$.
These pseudo-paired samples are subsequently used to train the conditional
flow-matching model.

\subsection{Conditional flow matching for perturbation response prediction}\label{method:fm}
Given the OT-guided pseudo-pair $(\bm{x}_i^u,\bm{x}_{j_i}^p)$ from Sec.~\ref{method:ot}, we utilize conditional flow matching to learn a perturbation-conditioned vector field that transports the source cell to its perturbed endpoint~\citep{lipman2023flow, tong2024improving}. 
For notational simplicity, we denote the pseudo-pair as $(\bm{x}^u,\bm{x}^p)$ in the following.
To model the trajectory, we define a linear interpolating path for each pair:
\begin{align}
\bm{x}_{t} = (1-t)\bm{x}^u + t\bm{x}^p, \qquad t\in[0,1],
\end{align}
which interpolates between the source cell $\bm{x}_{0}\!=\!\bm{x}^u$ and the perturbed target $\bm{x}_{1}\!=\!\bm{x}^p$.
The corresponding reference velocity can be defined by the following ordinary differential equation (ODE):
\begin{align}
v(\bm{x}_{t}) = \frac{d\bm{x}_{t}}{dt} = \bm{x}^p - \bm{x}^u,
\end{align}
This velocity can be approximated via a flow model $v_\theta(\bm{x}_{t}, t, \bm{\xi})$, where the perturbation vector $\bm{p}$ and biological context $\bm{c}$ serve as conditions, by minimizing the training objective:
\begin{align}
\mathcal{L}_{\mathrm{CFM}}(\theta)
=
\mathbb{E}_{(\bm{x}^u, \bm{x}^p,t,\bm{\xi})}
[
\|
v_\theta(\bm{x}_{t},t,\bm{\xi})
-
v(\bm{x}_{t})
\|^2
],
\end{align}
where $t$ is sampled from the uniform distribution on $[0,1]$.
At inference time, given an unperturbed cell $\bm{x}^u$ and a queried condition $(\bm{p},\bm{c})$, we initialize the trajectory at $\hat{\bm{x}}_0=\bm{x}^u$ and generate the perturbed state by integrating the learned vector field:
\begin{align}
\frac{d\hat{\bm{x}}_t}{dt}
=
v_\theta(\hat{\bm{x}}_t,t,\bm{\xi}),
\qquad t\in[0,1].
\end{align}
This ODE is solved numerically using Euler integration:
\begin{align}\label{fm:euler}
\hat{\bm{x}}_{t_{k+1}}
=
\hat{\bm{x}}_{t_k}
+
\Delta t\,
v_\theta(\hat{\bm{x}}_{t_k},t_k,\bm{\xi}),
\qquad
k=0,\ldots,K_{s}-1.
\end{align}
Here, $K_s$ is the number of uniform steps, with $\Delta t = 1/K_s$; we set $K_s=10$ for all experiments.

\subsection{Energy guidance for perturbation response refinement}

Predicting cellular perturbations has long suffered from poor generalization,
primarily due to the diverse spectrum of drug types and dosages, as well as
the combinatorial complexity of joint perturbations~\citep{hetzel2022chemcpa,wei2026benchmarking}.
Current models~\citep{hetzel2022chemcpa,lotfollahi2023cpa,klein2025cellflow}
typically use continuous vectors with biological or chemical structural
meaning as conditional embeddings, aiming to extrapolate learned patterns
to unseen perturbations.
In contrast, we alleviate this issue through a post-processing step termed
energy relaxation, which iteratively adjusts the predicted response to reduce
its condition-specific energy, thereby encouraging greater compatibility with
the perturbed distribution~\citep{du2019implicit}.

\textbf{Conditional energy guidance.}
Here, we define an energy function
$E_\phi(\bm{x}_t,\bm{\xi})$
that measures the compatibility between the condition
$\bm{\xi}$ and the intermediate state $\bm{x}_t$
interpolated between $\bm{x}^u$ and $\bm{x}^p$, where a lower
energy indicates stronger consistency with
$\bm{\xi}$.
Following the Boltzmann distribution, we assume that the conditional
probability
$p(\bm{\xi}|\bm{x}_t)$
is exponentially inversely related to the energy function
$E_\phi(\bm{x}_t,\bm{\xi})$
~\citep{duvenaud2020your,du2019implicit}:
\begin{align}
p(\bm{\xi}|\bm{x}_t)
~\propto~
\exp{
-\lambda_t
E_\phi}(\bm{x}_t,\bm{\xi})
,
\end{align}
where $\lambda_t>0$ controls the scale of the conditional energy and,
consequently, the strength of the resulting energy guidance.
Taking the gradient of the log-likelihood with respect to $\bm{x}_t$
yields:
\begin{align}
\nabla_{\bm{x}_t}
\log p(\bm{\xi}|\bm{x}_t)
~\propto~
-\lambda_t
\nabla_{\bm{x}_t}
E_\phi(\bm{x}_t,\bm{\xi}),
\end{align}
which is referred to as energy guidance.
Thus, increasing the conditional probability is mathematically equivalent
to incorporating energy guidance into the Euler integration step of flow
matching:
\begin{align}\label{energy_guidance}
\tilde{\bm{x}}_{t_{k+1}}
=
\hat{\bm{x}}_{t_{k+1}}
-
\lambda_t
\nabla_{\hat{\bm{x}}_{t_{k+1}}}
E_\phi(\hat{\bm{x}}_{t_{k+1}},\bm{\xi}),
\end{align}
where $\hat{\bm{x}}_{t_{k+1}}$ is given by Eq.~\ref{fm:euler},
and the same $\lambda_t$ controls the magnitude of the energy-guided
correction.
Under multiple independent conditions, the total energy in our method can
be approximated as a weighted sum of the energy functions for individual
condition components.
Specifically, we denote the independent condition components as
$\{\bm{\xi}_1,\ldots,\bm{\xi}_n\}$.
A product-of-experts approximation is used to combine the condition-specific distributions as follows:
\begin{align}
p(
\bm{\xi}_1,
\ldots,
\bm{\xi}_n
\mid
\bm{x}_t
)
~\propto~
\prod_{i=1}^{n}
p(
\bm{\xi}_i
\mid
\bm{x}_t
).
\end{align}
Accordingly, the corresponding total energy can be approximated as
\begin{align}
E_\phi
(
\bm{x}_t,
\{\bm{\xi}_1,\ldots,\bm{\xi}_n\}
)
\approx
\lambda_1
\mathcal{D}_{\boldsymbol{\phi}_1}
(
\bm{x}_t,\bm{\xi}_1
)
+
\cdots
+
\lambda_n
\mathcal{D}_{\boldsymbol{\phi}_n}
(
\bm{x}_t,\bm{\xi}_n
).
\end{align}
In practice, we employ a set of distinct distance functions
$\{
\mathcal{D}_{\boldsymbol{\phi}_1}(\cdot,\cdot),
\dots,
\mathcal{D}_{\boldsymbol{\phi}_n}(\cdot,\cdot)
\}$
tailored to specific condition components, such as drug identity, dosage,
or cellular context.

\textbf{Distance function construction.}
Based on the above analysis, the distance function should assign lower energy
to condition-consistent states and higher energy otherwise.
To this end, we parameterize the distance function with an energy model
$\mathcal{D}_{\boldsymbol{\phi}}$, which takes a cellular state and a condition
$\bm{\xi}$ as input and outputs a scalar energy score.
The energy model is trained separately from the flow model with the latter frozen.

For condition discrimination, we use measured perturbed cells as reference
states. Specifically, a measured perturbed cell $\bm{x}^{p}$ paired with its
original perturbation condition $\bm{\xi}^{+}$ forms a positive sample,
whereas a negative sample $(\bm{x}^{p},\bm{\xi}^{-})$ is constructed by
keeping the cellular state fixed and modifying the perturbation condition.
The condition-discrimination loss is defined as
\begin{align}
\mathcal{L}_{\mathrm{cond}}
=
\mathbb{E}_{
(\bm{x}^{p},\bm{\xi}^{+})\in\mathcal{P}^{+},
(\bm{x}^{p},\bm{\xi}^{-})\in\mathcal{P}^{-}
}
[
\operatorname{softplus}(
\mathcal{D}_{\boldsymbol{\phi}}(\bm{x}^{p},\bm{\xi}^{+})
-
\mathcal{D}_{\boldsymbol{\phi}}(\bm{x}^{p},\bm{\xi}^{-})
+
\Delta_{\mathrm{margin}}
)].
\end{align}
Here, $\mathcal{P}^{+}$ and $\mathcal{P}^{-}$ denote the positive and negative
sets, respectively, and $\Delta_{\mathrm{margin}}>0$ denotes the margin.
This objective encourages the measured response to receive lower energy under
its matched perturbation condition than under a mismatched condition.

Moreover, to encourage a monotonic decrease in energy along the transition 
toward the measured perturbed state, we introduce an additional path-ranking objective. 
Specifically, we first sample an intermediate state $\hat{\bm{x}}_{t}$ from the frozen flow trajectory and construct interpolated states between $\hat{\bm{x}}_{t}$ and the measured perturbed state $\bm{x}^{p}$ as
\begin{align}
\bm{x}_{\alpha}
=
(1-\alpha)\hat{\bm{x}}_{t}
+
\alpha\bm{x}^{p},
\qquad
\alpha\in\{0,0.25,0.5,0.75,1\}.
\end{align}
For two interpolated states $\bm{x}_{\alpha_1}$ and $\bm{x}_{\alpha_2}$ with
$\alpha_1<\alpha_2$, the path-ranking loss is defined as
\begin{align}
\mathcal{L}_{\mathrm{path}}
=
\mathbb{E}_{\alpha_1<\alpha_2}
[
\operatorname{softplus}
(
\mathcal{D}_{\boldsymbol{\phi}}
(\bm{x}_{\alpha_2},\bm{\xi})
-
\mathcal{D}_{\boldsymbol{\phi}}
(\bm{x}_{\alpha_1},\bm{\xi})
+
\Delta_{\mathrm{margin}}
)
],
\end{align}
where $\bm{x}_{\alpha_2}$ is encouraged to have lower energy than
$\bm{x}_{\alpha_1}$. The final objective for training the energy score model is defined as
\begin{align}
\mathcal{L}_{\mathrm{energy}}
=
\omega_{1}\mathcal{L}_{\mathrm{cond}}
+
\omega_{2}\mathcal{L}_{\mathrm{path}},
\end{align}
where $\omega_{1}=1$ and $\omega_{2}=1$ are weighting coefficients.

In our framework, we instantiate this objective for two energy models that capture drug and dose consistency.
For each model, negative samples are generated by perturbing only the corresponding condition while keeping the remaining conditions fixed:
$\mathcal{P}^{-}_{\mathrm{drug}}$ changes drug and
$\mathcal{P}^{-}_{\mathrm{dose}}$ changes dose.
Each model is trained using the objective above and its corresponding negative set.

\subsection{Datasets and preprocessing}\label{exp:dataset}

\textbf{4i dataset~\citep{bunne2023learning}}. 
Derived from two co-cultured primary melanoma cell lines (M130219 and M130429), this dataset tracks individual cellular responses to various chemical perturbations. It encompasses a total of $77,743$ cells evaluated across $27$ different drug treatments.
Each cell is characterized by $48$ distinct cellular attributes, comprising $22$ morphological features and $26$ protein intensities.

\textbf{SciPlex3~\citep{srivatsan2020massively}.} 
This benchmark captures transcriptional responses under extensive chemical stress across three human cancer cell lines (A549, K562, and MCF7). It catalogs $649,340$ single-cell transcriptomes exposed to $188$ small-molecule compounds, each screened at four escalating doses ($10\text{ nM}$, $100\text{ nM}$, $1\ \mu\text{M}$, and $10\ \mu\text{M}$) to resolve fine-grained, pathway-specific variations.
For computational efficiency, we subsample 290,888 cells and retain the top 2,000 highly variable genes (HVGs).

\textbf{ComboSciPlex~\citep{lotfollahi2023cpa}.} 
Focusing on the human lung adenocarcinoma cell line (A549), this dataset examines single-cell transcriptional profiles under single-drug and combination treatment regimens. The original cohort aggregates $63,378$ cells across $32$ conditions, including control, single-drug treatments, and drug combinations, from which the top 2,000 highly variable genes are selected for our experiments.

\textbf{LINCS L1000~\citep{subramanian2017next}.}
LINCS L1000 is a large-scale perturbational transcriptomics dataset that profiles gene-expression responses to chemical perturbations across diverse human cell lines, with 978 landmark genes. We focus on small-molecule perturbations measured at 24~h, resulting in 24,104 expression profiles across 21 cell lines and 79 compounds. 
L1000 data are provided at multiple processing levels; we use Level 3 normalized expression profiles for model training and prediction, while Level 5 consensus differential-expression signatures are used only as reference signatures to evaluate the directional concordance of predictions under held-out perturbation conditions.

\subsection{Evaluation metrics}\label{exp:metric}
Given a perturbation condition $\bm{p}$ within a biological context $\bm{c}$, let $\mathcal{B}^{\mathrm{u}}\!=\!\{\bm{x}^{u}_i\}_{i=1}^{N^{u}}$ denote the unperturbed control cells, and let $\mathcal{B}^{\mathrm{p}}\!=\!\{\bm{x}^{p}_j\}_{j=1}^{N^{p}}$ denote the measured perturbed cells. 
Following~\citep{bunne2023learning}, we subsample the unperturbed control and matched observed populations to an equal size of $N^{\prime}$ cells for computational efficiency, and denote the predicted perturbed cells as $\hat{\mathcal{B}}^{\mathrm{p}} = \{\hat{\bm{x}}^{p}_k\}_{k=1}^{N^{\prime}}$. 
Here, $N^{\prime}$ is selected as the maximum possible value from $\{100, 250, 500, 1000, 1500\}$ given the available cell counts, satisfying $N^{\prime}\!\le\!\min(N^{u}, N^{p})$.
Notably, all downsampling validations are repeated 10 times with the mean results reported to ensure robustness.

\textbf{Pearson correlation coefficient.}
This metric measures agreement between predicted and measured population-level mean profiles across features and is defined as follows:
\begin{align}
\mathrm{Pearson}
=
\frac{
\sum_{k=1}^{D}
(\bar{x}^{p}_{k}-\mu^{p})
(\bar{\hat{x}}^{p}_{k}-\hat{\mu}^{p})
}
{
\sqrt{
\sum_{k=1}^{D}
(\bar{x}^{p}_{k}-\mu^{p})^2
}
\cdot
\sqrt{
\sum_{k=1}^{D}
(\bar{\hat{x}}^{p}_{k}-\hat{\mu}^{p})^2
}
},
\end{align}
where $D$ is the dimension of features, $\bar{\bm{x}}^{p}$ and $\bar{\hat{\bm{x}}}^{p}$ denote the measured and predicted values averaged over samples, respectively, while $\mu^{p}$ and $\hat{\mu}^{p}$ represent their means across all features.

\textbf{Squared maximum mean discrepancy (MMD$^2$).}
MMD$^2$ is a kernel-based metric for comparing two probability distributions~\citep{gretton2012kernel}. 
Unlike the Pearson correlation coefficient, which evaluates the agreement between mean response profiles, it compares entire distributions via pairwise sample similarities, thereby capturing discrepancies in higher-order population structures.
Formally, MMD$^2$ is defined as:
\begin{align}
\mathrm{MMD}^{2}(\mathcal{B}^{\mathrm{p}},\hat{\mathcal{B}}^{\mathrm{p}})
=
\mathbb{E}[k_{\mathrm{ms}}(\bm{x}^{p},\bm{x}^{p^\prime})]
+
\mathbb{E}[k_{\mathrm{ms}}(\hat{\bm{x}}^{p},\hat{\bm{x}}^{p^\prime})]
-
2\mathbb{E}[k_{\mathrm{ms}}(\bm{x}^{p},\hat{\bm{x}}^{p})].
\end{align}
Here, $k_{\mathrm{ms}}$ is a multi-scale Gaussian radial basis function (RBF) kernel designed to capture discrepancies at different resolutions (from local cell-level variations to global population-level shifts):
\begin{align}
k_{\mathrm{ms}}(\bm{x}^{p},\hat{\bm{x}}^{p})
=
\frac{1}{L}\sum_{\ell=1}^{L}
\exp(-\gamma_{\ell}\|\bm{x}^{p}-\hat{\bm{x}}^{p}\|^{2}),
\end{align}
where $L=50$ and $\gamma_\ell$ is logarithmically spaced from $10^{1}$ to $10^{-3}$.

\textbf{Energy distance~\citep{szekely2013energy}.}
This criterion measures the statistical distance between predicted and observed cell population attributes, which characterizes the equality of their distributions.
It reflects the extent to which the predicted cells capture the overall geometry, dispersion, and variability of the observed perturbed population.
Formally, the distance is defined as:
\begin{align}
\text{ED}(\mathcal{B}^{\mathrm{p}},\hat{\mathcal{B}}^{\mathrm{p}}) = 2\mathbb{E}\|\bm{x}^{p} - \hat{\bm{x}}^{p}\| - \mathbb{E}\|\bm{x}^{p} - \bm{x}^{p^\prime}\| - \mathbb{E}\|\hat{\bm{x}}^{p} - \hat{\bm{x}}^{p^\prime}\|,
\end{align}
where $\|\cdot\|$ is the pairwise Euclidean distance. 
Here, $\bm{x}^{p},\bm{x}^{p^\prime}$ are sampled uniformly and independently from the observed cell population $\mathcal{B}^{\mathrm{p}}$, and $\hat{\bm{x}}^{p},\hat{\bm{x}}^{p^\prime}$ are drawn from the predicted cell population $\hat{\mathcal{B}}^{\mathrm{p}}$.

\textbf{Sinkhorn divergence~\citep{feydy2019interpolating}.}
To assess the geometric alignment between the observed cell population $\mathcal{B}^{p}$ and the predicted population $\hat{\mathcal{B}}^{p}$, we utilize the Sinkhorn divergence $S_{\varepsilon}$, defined as:
\begin{align}
S_{\varepsilon}(\mathcal{B}^{p}, \hat{\mathcal{B}}^{p}) = \text{OT}_{\varepsilon}(\mathcal{B}^{p}, \hat{\mathcal{B}}^{p}) - \frac{1}{2}\text{OT}_{\varepsilon}(\mathcal{B}^{p}, \mathcal{B}^{p}) - \frac{1}{2}\text{OT}_{\varepsilon}(\hat{\mathcal{B}}^{p}, \hat{\mathcal{B}}^{p}),
\end{align}
where $\text{OT}_{\varepsilon}$ is the entropy-regularized optimal transport cost with a regularization strength $\varepsilon$.
Specifically, given the evaluation cost matrix 
$\bm{C}^{\mathrm{eval}}\in\mathbb{R}^{N^{\prime}\times N^{\prime}}$ with 
$C^{\mathrm{eval}}_{ij}=\|\bm{x}^{p}_i-\hat{\bm{x}}^{p}_j\|_2^2$, the entropic OT cost used for evaluation is formulated as
\begin{align}
\mathrm{OT}_{\varepsilon}(\mathcal{B}^{p}, \hat{\mathcal{B}}^{p}) 
= 
\min_{\bm{T}\in\Pi(\bm{u}^{\mathrm{eval}},\bm{p}^{\mathrm{eval}})}
\sum_{i,j}T_{ij}C^{\mathrm{eval}}_{ij}
+
\varepsilon
\sum_{i,j}T_{ij}(\log T_{ij}-1),
\end{align}
where $\Pi(\bm{u}^{\mathrm{eval}},\bm{p}^{\mathrm{eval}})$ is defined analogously to Eq.~\ref{eq:ot_sub}, with uniform empirical weights over the observed and predicted cells. Here, $\bm{T}$ denotes the transport plan between the two cell populations, and $\varepsilon$ is the entropic regularization strength. 
In practice, each $\mathrm{OT}_{\varepsilon}$ term is computed using the Sinkhorn algorithm. For evaluation, we set $\varepsilon=100$ and use $100$ Sinkhorn iterations.

\subsection{Experimental setup}\label{sec11_subsec4}


\textbf{Evaluation settings.}
We evaluate models under three settings:
\textbf{1)} In the independent and identically distributed (IID) setting, each perturbation condition appears in training, with cells split into 80\% training and 20\% test cells within each condition. 
This setting measures interpolation within the observed perturbation space.
\textbf{2)} In the out-of-distribution (OOD) setting, the perturbation conditions used for testing are absent from the training set. For SciPlex3~\citep{srivatsan2020massively}, we hold out three drug-dose conditions, and for ComboSciPlex~\citep{lotfollahi2023cpa}, we hold out five drug-combination conditions. This setting tests whether models can generalize to unseen dosage or combination conditions rather than interpolate among observed conditions. 
\textbf{3)} In the few-shot setting, we evaluate a related but less stringent scenario in which the target perturbations are no longer completely unseen: only a small number of cells from each target condition is included during training, and prediction is evaluated on the remaining held-out cells from the same conditions.
We use this setting on the 4i~\citep{bunne2023learning} dataset for five target perturbations under a leave-one-condition protocol. 
This setting tests adaptation when only limited condition-specific supervision is available. See Appendix for more details.

\textbf{Baselines.}
Our method is benchmarked against leading single-cell perturbation prediction models, including CellOT~\citep{bunne2023learning}, biolord~\citep{piran2024disentanglement}, CPA~\citep{lotfollahi2023cpa}, PRNet~\citep{qi2024prnet}, CellFlow~\citep{klein2025cellflow}, and scDFM~\citep{yu2026scdfm}.
We also include \textit{MeanExp} as a statistical baseline~\citep{wei2026benchmarking}. 
In the IID setting, \textit{MeanExp} predicts the mean expression profile of training samples from the queried condition.
In the OOD setting, it predicts the mean expression of all perturbed training cells with the same dosage as the target condition. 
The official implementations of all baseline methods are used with their recommended hyperparameters and dataset-specific adjustments. 
For a fair comparison, all models are evaluated using the same data protocol.

\textbf{Implementation details.}
Our model is implemented in PyTorch~\citep{paszke2019pytorch} and optimized with AdamW using a batch size of 256 for 4i, SciPlex3, and ComboSciPlex. 
For \textbf{4i}, we train the model for 100,000 iterations with a learning rate of \(1\times10^{-6}\), Sinkhorn regularization \(\varepsilon_{\mathrm{OT}}=0.2\), and 100 Sinkhorn iterations.
For \textbf{SciPlex3}~\citep{srivatsan2020massively}, we use 50,000 training iterations, a learning rate of $1\times 10^{-7}$, $\varepsilon_{\mathrm{OT}}=0.1$, and 100 Sinkhorn iterations.
\textbf{ComboSciPlex}~\citep{lotfollahi2023cpa} uses the same optimization and mini-batch OT settings as SciPlex3.
For LINCS L1000~\citep{subramanian2017next}, we use a batch size of 128, train the conditional flow-matching model for 30,000 iterations with a learning rate of \(1\times10^{-7}\), and train the energy model for 300 iterations with a learning rate of \(1\times10^{-4}\).
During inference, energy-guided refinement is performed for 10 steps with guidance coefficient $\lambda=0.05$.

\bibliographystyle{unsrt}
\bibliography{references}

@String(ICML  = {ICML})

@String(NIPS  = {NeurIPS})

@String(ICLR  = {ICLR})

@article{bunne2023learning,
  title={Learning single-cell perturbation responses using neural optimal transport},
  author={Bunne, Charlotte and Stark, Stefan G and Gut, Gabriele and Del Castillo, Jacobo Sarabia and Levesque, Mitch and Lehmann, Kjong-Van and Pelkmans, Lucas and Krause, Andreas and R{\"a}tsch, Gunnar},
  journal={Nature Methods},
  year={2023}
}

@article{lotfollahi2019scgen,
    title   = {scGen predicts single-cell perturbation responses},
    author  = {Lotfollahi, Mohammad and Wolf, F Alexander and Theis, Fabian J},
    journal = {Nature Methods},
    year    = {2019},
}

@article{dixit2016perturbseq,
  title={Perturb-Seq: Dissecting Molecular Circuits with Scalable Single-Cell RNA Profiling of Pooled Genetic Screens},
  author={Dixit, Atray and Parnas, Oren and Li, Biyu and Chen, Jenny and Fulco, Charles P. and Jerby-Arnon, Livnat and Marjanovic, Nemanja D. and Dionne, Danielle and Burks, Tyler and Raychowdhury, Raktima and Adamson, Britt and Norman, Thomas M. and Lander, Eric S. and Weissman, Jonathan S. and Friedman, Nir and Regev, Aviv},
  journal={Cell},
  year={2016}
}

@article{norman2019exploring,
  title={Exploring genetic interaction manifolds constructed from rich single-cell phenotypes},
  author={Norman, Thomas M. and Horlbeck, Max A. and Replogle, Joseph M. and Ge, Alex Y. and Xu, Albert and Jost, Marco and Gilbert, Luke A. and Weissman, Jonathan S.},
  journal={Science},
  year={2019}
}

@article{srivatsan2020massively,
  title={Massively multiplex chemical transcriptomics at single-cell resolution},
  author={Srivatsan, Sanjay R. and McFaline-Figueroa, Jos{\'e} L. and Ramani, Vijay and Saunders, Lauren and Cao, Junyue and Packer, Jonathan and Pliner, Hannah A. and Jackson, Dana L. and Daza, Riza M. and Christiansen, Lena and Zhang, Fan and Steemers, Frank and Shendure, Jay and Trapnell, Cole},
  journal={Science},
  year={2020}
}

@article{kang2018multiplexed,
  title={Multiplexed droplet single-cell {RNA}-sequencing using natural genetic variation},
  author={Kang, Hyun Min and Subramaniam, Meena and Targ, Sasha and Nguyen, Michelle and Maliskova, Lenka and McCarthy, Elizabeth and Wan, Eunice and Wong, Simon and Byrnes, Lauren and Lanata, Cristina M. and Gate, Rachel E. and Mostafavi, Sara and Marson, Alexander and Zaitlen, Noah and Criswell, Lindsey A. and Ye, Chun Jimmie},
  journal={Nature Biotechnology},
  year={2018}
}

@article{replogle2022mapping,
  title={Mapping information-rich genotype-phenotype landscapes with genome-scale Perturb-seq},
  author={Replogle, Joseph M. and Saunders, Reuben A. and Pogson, Angela N. and Hussmann, Jeffrey A. and Lenail, Alexander and Guna, Alina and Mascibroda, Lauren and Wagner, Eric J. and Adelman, Karen and Lithwick-Yanai, Gila and Iremadze, Nika and Oberstrass, Florian and Lipson, Doron and Bonnar, Jessica L. and Jost, Marco and Norman, Thomas M. and Weissman, Jonathan S.},
  journal={Cell},
  year={2022}
}

@article{schiebinger2019optimal,
  title={Optimal-Transport Analysis of Single-Cell Gene Expression Identifies Developmental Trajectories in Reprogramming},
  author={Schiebinger, Geoffrey and Shu, Jian and Tabaka, Marcin and Cleary, Brian and Subramanian, Vidya and Solomon, Aryeh and Gould, Joshua and Liu, Siyan and Lin, Stacie and Berube, Peter and Lee, Lia and Chen, Jenny and Brumbaugh, Justin and Rigollet, Philippe and Hochedlinger, Konrad and Jaenisch, Rudolf and Regev, Aviv and Lander, Eric S.},
  journal={Cell},
  year={2019}
}

@inproceedings{tong2020trajectorynet,
  title={TrajectoryNet: A Dynamic Optimal Transport Network for Modeling Cellular Dynamics},
  author={Tong, Alexander and Huang, Jessie and Wolf, Guy and Van Dijk, David and Krishnaswamy, Smita},
  booktitle=ICML,
  year={2020}
}

@inproceedings{cuturi2013sinkhorn,
  title={Sinkhorn Distances: Lightspeed Computation of Optimal Transport},
  author={Cuturi, Marco},
  booktitle=NIPS,
  year={2013}
}

@inproceedings{lipman2023flow,
  title={Flow Matching for Generative Modeling},
  author={Lipman, Yaron and Chen, Ricky T. Q. and Ben-Hamu, Heli and Nickel, Maximilian and Le, Matthew},
  booktitle=ICLR,
  year={2023}
}

@article{lotfollahi2023cpa,
  title={Predicting cellular responses to complex perturbations in high-throughput screens},
  author={Lotfollahi, Mohammad and Klimovskaia Susmelj, Anna and De Donno, Carlo and Hetzel, Leon and Ji, Yuge and Ibarra, Ignacio L. and Srivatsan, Sanjay R. and Naghipourfar, Mohsen and Daza, Riza M. and Martin, Beth and Shendure, Jay and McFaline-Figueroa, Jose L. and Boyeau, Pierre and Wolf, F. Alexander and Yakubova, Nafissa and G{\"u}nnemann, Stephan and Trapnell, Cole and Lopez-Paz, David and Theis, Fabian J.},
  journal={Molecular Systems Biology},
  year={2023}
}

@inproceedings{hetzel2022chemcpa,
  title={Predicting Cellular Responses to Novel Drug Perturbations at a Single-Cell Resolution},
  author={Hetzel, Leon and Boehm, Simon and Kilbertus, Niki and G{\"u}nnemann, Stephan and Lotfollahi, Mohammad and Theis, Fabian J.},
  booktitle=NIPS,
  year={2022}
}

@article{qi2024prnet,
  title={Predicting transcriptional responses to novel chemical perturbations using deep generative model for drug discovery},
  author={Qi, Xiaoning and Zhao, Lianhe and Tian, Chenyu and Li, Yueyue and Chen, Zhen-Lin and Huo, Peipei and Chen, Runsheng and Liu, Xiaodong and Wan, Baoping and Yang, Shengyong and Zhao, Yi},
  journal={Nature Communications},
  year={2024}
}

@article{roohani2024gears,
  title={Predicting transcriptional outcomes of novel multigene perturbations with {GEARS}},
  author={Roohani, Yusuf and Huang, Kexin and Leskovec, Jure},
  journal={Nature Biotechnology},
  year={2024}
}

@article{klein2025cellflow,
  title={{CellFlow} enables generative single-cell phenotype modeling with flow matching},
  author={Klein, Dominik and Fleck, Jonas Simon and Bobrovskiy, Daniil and Zimmermann, Lea and Becker, S{\"o}ren and Palma, Alessandro and Dony, Leander and Tejada-Lapuerta, Alejandro and Huguet, Guillaume and Lin, Hsiu-Chuan and Azbukina, Nadezhda and Sanch{\'i}s-Calleja, F{\'a}tima and Uscidda, Theo and Szalata, Artur and Gander, Manuel and Regev, Aviv and Treutlein, Barbara and Camp, J. Gray and Theis, Fabian J.},
  journal={bioRxiv},
  year={2025}
}

@article{vinas2025systema,
  title={{Systema}: a framework for evaluating genetic perturbation response prediction beyond systematic variation},
  author={Vi{\~n}as Torn{\'e}, Ramon and Wiatrak, Maciej and Piran, Zoe and Fan, Shuyang and Jiang, Liangze and Teichmann, Sarah A. and Nitzan, Mor and Brbi{\'c}, Maria},
  journal={Nature Biotechnology},
  year={2025},
  doi={10.1038/s41587-025-02777-8}
}

@article{csendes2025benchmarking,
  title={Benchmarking foundation cell models for post-perturbation {RNA}-seq prediction},
  author={Csendes, Gerold and Sanz, Gema and Szalay, Krist{\'o}f Z. and Szalai, Bence},
  journal={BMC Genomics},
  year={2025}
}

@article{tong2024improving,
  title={Improving and generalizing flow-based generative models with minibatch optimal transport},
  author={Tong, Alexander and Fatras, Kilian and Malkin, Nikolay and Huguet, Guillaume and Zhang, Yanlei and Rector-Brooks, Jarrid and Wolf, Guy and Bengio, Yoshua},
  journal={Transactions on Machine Learning Research},
  year={2024}
}

@inproceedings{paszke2019pytorch,
  title={{PyTorch}: An Imperative Style, High-Performance Deep Learning Library},
  author={Paszke, Adam and Gross, Sam and Massa, Francisco and Lerer, Adam and Bradbury, James and Chanan, Gregory and Killeen, Trevor and Lin, Zeming and Gimelshein, Natalia and Antiga, Luca and Desmaison, Alban and Kopf, Andreas and Yang, Edward and DeVito, Zachary and Raison, Martin and Tejani, Alykhan and Chilamkurthy, Sasank and Steiner, Benoit and Fang, Lu and Bai, Junjie and Chintala, Soumith},
  booktitle=NIPS,
  year={2019}
}

@article{szekely2013energy,
  title={Energy statistics: A class of statistics based on distances},
  author={Sz{\'e}kely, G{\'a}bor J. and Rizzo, Maria L.},
  journal={Journal of Statistical Planning and Inference},
  year={2013}
}

@inproceedings{feydy2019interpolating,
  title={Interpolating between Optimal Transport and {MMD} using Sinkhorn Divergences},
  author={Feydy, Jean and S{\'e}journ{\'e}, Thibault and Vialard, Fran{\c{c}}ois-Xavier and Amari, Shun-ichi and Trouv{\'e}, Alain and Peyr{\'e}, Gabriel},
  booktitle={Proceedings of the Twenty-Second International Conference on Artificial Intelligence and Statistics},
  year={2019}
}

@article{gretton2012kernel,
  title={A Kernel Two-Sample Test},
  author={Gretton, Arthur and Borgwardt, Karsten M. and Rasch, Malte J. and Sch{\"o}lkopf, Bernhard and Smola, Alexander},
  journal={Journal of Machine Learning Research},
  year={2012}
}

@inproceedings{yu2026scdfm,
  title={sc{DFM}: Distributional Flow Matching Model for Robust Single-Cell Perturbation Prediction},
  author={Chenglei Yu and Chuanrui Wang and Bangyan Liao and Tailin Wu},
  booktitle=ICLR,
  year={2026}
}

@article{piran2024disentanglement,
  title={Disentanglement of single-cell data with biolord},
  author={Piran, Zoe and Cohen, Niv and Hoshen, Yedid and Nitzan, Mor},
  journal={Nature Biotechnology},
  year={2024}
}

@article{wei2026benchmarking,
  title={Benchmarking algorithms for generalizable single-cell perturbation response prediction},
  author={Wei, Zhiting and Wang, Yiheng and Gao, Yicheng and Wang, Shuguang and Li, Ping and Si, Duanmiao and Gao, Yuli and Wu, Siqi and Li, Danlu and Dong, Kejing and others},
  journal={Nature Methods},
  year={2026}
}

@article{duvenaud2020your,
  title={Your Classifier is Secretly an Energy Based Model and You Should Treat it Like One},
  author={Grathwohl, Will and Wang, Kuan-Chieh and Jacobsen, J{\"o}rn-Henrik and Duvenaud, David and Norouzi, Mohammad and Swersky, Kevin},
  journal=ICLR,
  year={2020}
}

@article{du2019implicit,
  title={Implicit Generation and Modeling in Energy-Based Models},
  author={Du, Yilun and Mordatch, Igor},
  journal=NIPS,
  year={2019}
}

@article{ahlmann2025deep,
  title={Deep-learning-based gene perturbation effect prediction does not yet outperform simple linear baselines},
  author={Ahlmann-Eltze, Constantin and Huber, Wolfgang and Anders, Simon},
  journal={Nature Methods},
  year={2025}
}

@inproceedings{kim2023swamp,
  title={SwAMP: Swapped Assignment of Multi-Modal Pairs for Cross-Modal Retrieval},
  author={Kim, Minyoung},
  booktitle={Proceedings of the 26th International Conference on Artificial Intelligence and Statistics},
  year={2023}
}

@article{cui2024scgpt,
  title={scGPT: toward building a foundation model for single-cell multi-omics using generative AI},
  author={Cui, Haotian and Wang, Chloe and Maan, Hassaan and Pang, Kuan and Luo, Fengning and Duan, Nan and Wang, Bo},
  journal={Nature Methods},
  year={2024}
}

@article{hao2024large,
  title={Large-scale foundation model on single-cell transcriptomics},
  author={Hao, Minsheng and Gong, Jing and Zeng, Xin and Liu, Chiming and Guo, Yucheng and Cheng, Xingyi and Wang, Taifeng and Ma, Jianzhu and Zhang, Xuegong and Song, Le},
  journal={Nature Methods},
  year={2024}
}

@article{mcfarland2020multiplexed,
  title={Multiplexed single-cell transcriptional response profiling to define cancer vulnerabilities and therapeutic mechanism of action},
  author={McFarland, James M and Paolella, Brenton R and Warren, Allison and Geiger-Schuller, Kathryn and Shibue, Tsukasa and Rothberg, Michael and Kuksenko, Olena and Colgan, William N and Jones, Andrew and Chambers, Emily and others},
  journal={Nature communications},
  year={2020}
}

@article{palmer2017combination,
  title={Combination cancer therapy can confer benefit via patient-to-patient variability without drug additivity or synergy},
  author={Palmer, Adam C and Sorger, Peter K},
  journal={Cell},
  year={2017}
}

@article{lamb2006connectivity,
  title={The Connectivity Map: using gene-expression signatures to connect small molecules, genes, and disease},
  author={Lamb, Justin and Crawford, Emily D and Peck, David and Modell, Joshua W and Blat, Irene C and Wrobel, Matthew J and Lerner, Jim and Brunet, Jean-Philippe and Subramanian, Aravind and Ross, Kenneth N and others},
  journal={science},
  year={2006}
}

@article{subramanian2017next,
  title={A next generation connectivity map: L1000 platform and the first 1,000,000 profiles},
  author={Subramanian, Aravind and Narayan, Rajiv and Corsello, Steven M and Peck, David D and Natoli, Ted E and Lu, Xiaodong and Gould, Joshua and Davis, John F and Tubelli, Andrew A and Asiedu, Jacob K and others},
  journal={Cell},
  year={2017}
}

@article{peng2026generative,
  title={A generative framework for predicting cellular morphological and transcriptomic perturbation responses},
  author={Peng, Rui and Liu, Ziru and Gao, Yuanxu and Wang, Jinzhuo},
  journal={Cell Reports Methods},
  year={2026}
}

\clearpage
\section*{Supplementary Information}
\addcontentsline{toc}{section}{Supplementary Information}

\setcounter{section}{0}
\setcounter{subsection}{0}
\setcounter{figure}{0}
\setcounter{table}{0}
\setcounter{equation}{0}
\renewcommand{\thesection}{S\arabic{section}}
\renewcommand{\thesubsection}{\thesection.\arabic{subsection}}
\renewcommand{\thefigure}{S\arabic{figure}}
\renewcommand{\thetable}{S\arabic{table}}
\renewcommand{\theequation}{S\arabic{equation}}
\section{Datasets and Preprocessing}
\label{sec:supp_datasets}

\subsection{Dataset overview}
\label{sec:supp_dataset_overview}

In this study, three single-cell perturbation datasets are used: 4i~\citep{bunne2023learning} provides imaging-based phenotypic measurements, SciPlex3~\citep{srivatsan2020massively} profiles transcriptional responses to single-drug treatments, and ComboSciPlex~\citep{lotfollahi2023cpa} profiles transcriptional responses to both single-drug and combination treatments.
For LINCS L1000~\citep{subramanian2017next}, Level 3 expression profiles are used for model training and prediction, and Level 5 consensus signatures are used to evaluate the direction of predicted gene-expression changes.
Table~\ref{tab:supp_dataset_overview} summarizes the processed datasets used for experiments.

\begin{table}[h]
\centering
\caption{Overview of the datasets used in this study.}
\label{tab:supp_dataset_overview}
\small
\setlength{\tabcolsep}{3.5pt}
\begin{tabular}{rccccc}
\toprule
Dataset & Cells/profiles & Features & Contexts & Conditions & Readout \\
\midrule
4i~\citep{bunne2023learning} & 77,743 & 48 & 1 & 27 & imaging phenotype \\
SciPlex3~\citep{srivatsan2020massively} & 290,888 & 2,000 & 3 & 749 & scRNA-seq \\
ComboSciPlex~\citep{lotfollahi2023cpa} & 63,378 & 2,000 & 1 & 32 & scRNA-seq \\
L1000 Level 3~\citep{subramanian2017next} & 24,104 & 978 & 21 & 110 & 
gene-expression profiles \\
\bottomrule
\end{tabular}
\end{table}

\subsection{4i preprocessing}
\label{sec:supp_4i_preprocessing}

The 4i dataset~\citep{bunne2023learning} contains single-cell imaging profiles under chemical perturbations. 
Following CellOT~\citep{bunne2023learning}, the preprocessed representation is retained without additional feature
filtering, with each cell characterized by 48 morphological and protein-marker intensity measurements.
The dataset contains 77,743 cells, including 10,995 untreated cells and 66,748 treated cells across 27 chemical perturbations.
Untreated cells serve as the source population, and cells treated with each compound form the corresponding target population.

For the held-out and few-shot experiments, five drugs are designated as target perturbations: \texttt{cisplatin}, \texttt{crizotinib}, \texttt{dabrafenib}, \texttt{dacarbazine}, and \texttt{dasatinib}.
Treated cells from these conditions are either excluded from training or included in limited numbers, following the evaluation protocols in Supplementary Section~\ref{sec:supp_protocols}.

\subsection{SciPlex3 preprocessing}
\label{sec:supp_sciplex_preprocessing}

SciPlex3~\citep{srivatsan2020massively} profiles transcriptional responses to chemical perturbations across three cancer cell lines:
A549, K562, and MCF7. 
Each cell line defines a distinct cellular context, with vehicle-treated cells serving as its source population.
Perturbation conditions are specified by compound identity and dose.
Data splits are defined before feature selection. 
The three drug-dose conditions, \ie, ENMD-2076 at 0.1~\(\mu\)M, WP1066 at 0.01~\(\mu\)M, and Mesna at 0.1~\(\mu\)M, are entirely withheld from training for OOD evaluation.
Gene-expression counts are normalized and log-transformed.
The top 2,000 highly variable genes are selected from the training cells using the Seurat procedure implemented in Scanpy.
The same gene set is retained for the IID and OOD test sets.
The processed dataset contains 290,888 cells represented by 2,000 genes across the three cellular contexts.
For scEGFlow training, these expression profiles are projected onto 200 principal components.
The PCA transformation is fitted exclusively on the training cells and then applied unchanged to the IID and OOD test sets.
The model predicts perturbed states in this PCA space, and these predictions are subsequently transformed back to the 2,000-gene expression space for gene-level evaluation.

\subsection{ComboSciPlex preprocessing}
\label{sec:supp_combo_preprocessing}
ComboSciPlex~\citep{lotfollahi2023cpa} profiles single-cell transcriptional responses to individual compounds and pairwise drug combinations in A549 cells.
This cell line defines a single cellular context, with cells receiving no active compound serving as the source population.
Perturbation conditions are specified by the individual compound or compound pair.
Data splits are defined before feature selection.
The five drug-combination conditions, i.e., \texttt{Dacinostat \!+\! Danusertib}, \texttt{Givinostat \!+\! Cediranib}, \texttt{Panobinostat \!+\! Alvespimycin}, \texttt{Panobinostat \!+\! SRT2104}, and \texttt{SRT2104 \!+\! Alvespimycin}, are entirely withheld from training for OOD evaluation.
Within each remaining observed perturbation condition, cells are partitioned into training and IID test sets.

Gene-expression counts are normalized and log-transformed following the same procedure as for SciPlex3~\citep{srivatsan2020massively} (Supplementary Section~\ref{sec:supp_sciplex_preprocessing}).
The top 2,000 highly variable genes are selected from the training cells using the Seurat procedure implemented in Scanpy.
The same gene set is retained for the IID and OOD test sets.
The processed dataset contains 63,378 cells represented by 2,000 genes across 32 conditions, including control, within a single cellular context.
For scEGFlow training, these expression profiles are projected onto 200 principal components.
The PCA transformation is fitted exclusively on the training cells and then applied unchanged to the IID and OOD test sets.
The model predicts perturbed states in this PCA space, and these predictions are subsequently transformed back to the 2,000-gene expression space for gene-level evaluation.

\subsection{LINCS L1000 preprocessing and external validation}
\label{sec:supp_l1000}

The LINCS L1000 compendium~\citep{subramanian2017next} profiles transcriptional responses to perturbations across multiple cell lines.
We restrict the analysis to compound treatments measured 24~h after exposure and their corresponding controls.
Each cell line defines a distinct cellular context, with untreated or vehicle-treated profiles serving as its source population.
Perturbation conditions are specified by compound identity and dose.
Data splits are defined before dimensionality reduction.
The six drug-dose conditions, i.e., \texttt{olaparib}, \texttt{GSK-1059615}, \texttt{dasatinib}, \texttt{PD-0325901}, \texttt{CGP-60474}, and \texttt{veliparib}, all at 10~\(\mu\mathrm{M}\), are entirely withheld from training for OOD evaluation.

Level~3 profiles provide normalized gene-expression measurements for individual experimental instances.
We retain the 978 directly measured landmark genes as the expression space.
The processed dataset contains 24,104 profiles represented by 978 genes across 21 cellular contexts and 79 compounds, including 13,672 control profiles and 10,432 compound-treated profiles.
For scEGFlow training, these expression profiles are projected onto 768 principal components.
The PCA transformation is fitted exclusively on the training profiles and then applied unchanged to held-out profiles.
The model predicts perturbed states in this PCA space, and these predictions are subsequently transformed back to the 978-gene
expression space for gene-level evaluation.

Level~5 consensus differential-expression signatures, represented as moderated z-scores, provide the reference directions of gene regulation for external validation.
These signatures are used only for evaluation and responsive-gene selection; they are not provided to either the scEGFlow backbone or the energy model during training or prediction.
For each held-out perturbation, scEGFlow generates predicted treated profiles from the corresponding Level~3 control population.
Predictions are matched to Level~5 signatures by cell line, compound identity, dose, and treatment duration.
The six held-out perturbations across 21 cell lines define 126 possible cell line-drug-dose settings.
Settings lacking the Level~3 profiles required for prediction or a corresponding Level~5 signature are excluded, leaving 111 matched settings for external validation.

We evaluate whether the model recovers the directions of gene regulation recorded in the Level~5 reference signatures.
For each matched cell line-drug-dose setting \(c\), we first compute the predicted expression change relative to the matched Level~3 controls.
We then select genes with the largest absolute Level~5 response scores and calculate the fraction for which the predicted and reference changes have the same direction.

For each landmark gene \(g\), the predicted expression change is the mean expression in the predicted treated population minus
the mean expression in the matched control population:
\begin{equation}
\Delta^{\mathrm{pred}}_{c,g}
=
\frac{1}{N^{\mathrm{pred}}_c}
\sum_{i=1}^{N^{\mathrm{pred}}_c}
\hat{x}^{\,p}_{c,i,g}
-
\frac{1}{N^{\mathrm{ctrl}}_c}
\sum_{j=1}^{N^{\mathrm{ctrl}}_c}
x^{\,u}_{c,j,g},
\end{equation}
where \(\hat{x}^{\,p}_{c,i,g}\) is the expression of gene \(g\) in predicted treated profile \(i\), and \(x^{\,u}_{c,j,g}\)
is its expression in matched Level~3 control profile \(j\).
Here, \(N^{\mathrm{pred}}_c\) and \(N^{\mathrm{ctrl}}_c\) denote the numbers of predicted treated and control profiles, respectively.
A positive value of \(\Delta^{\mathrm{pred}}_{c,g}\) indicates predicted upregulation, whereas a negative value indicates
predicted downregulation.

The Level~5 moderated z-score, denoted by \(\Delta^{\mathrm{L5}}_{c,g}\), provides the reference response
for gene \(g\) in setting \(c\), with its sign indicating the direction of regulation.
Because predicted expression differences and Level~5 scores are on different scales, we compare their signs rather than
their numerical magnitudes.
To evaluate direction agreement among genes with stronger reference responses, we rank the 978 landmark genes within each setting by \(|\Delta^{\mathrm{L5}}_{c,g}|\).
The top 10\%, approximately 98 genes per setting, form the responsive-gene set \(G_c\).
This selection uses only the Level~5 reference scores and is performed separately for each setting.

For the selected genes, directional concordance is the fraction whose predicted expression changes have the same sign as
their Level~5 reference scores:
\begin{equation}
\mathrm{Concordance}(c)
=
\frac{1}{|G_c|}
\sum_{g\in G_c}
\mathbf{1}
[
\operatorname{sign}(\Delta^{\mathrm{pred}}_{c,g})
=
\operatorname{sign}(\Delta^{\mathrm{L5}}_{c,g})
],
\end{equation}
where \(\mathbf{1}[\cdot]\) equals 1 when the signs agree and 0 otherwise.
Concordance is calculated separately for each of the 111 matched settings, with higher values indicating greater agreement
in the directions of gene regulation.

\section{Data Splits and Evaluation Protocols}
\label{sec:supp_protocols}

\subsection{IID experimental setup}
\label{sec:supp_iid}

For both 4i~\citep{bunne2023learning} and SciPlex3~\citep{srivatsan2020massively}, we randomly split the cells
within each perturbation condition into a training set (80\%) and a held-out test set (20\%). 
Every evaluated perturbation condition is therefore represented during training, but the individual test cells are excluded from training.

\subsection{OOD experimental setup}
\label{sec:supp_ood}

For OOD evaluation, treated samples from the target conditions are excluded from training. 
We consider three settings: unseen drug-dose conditions, transfer to an unseen cellular context, and unseen drug combinations.

\subsubsection{Drug-dose conditions}

For SciPlex3~\citep{srivatsan2020massively}, we exclude all treated cells from three drug-dose conditions from training: ENMD-2076 at 0.1~\(\mu\)M,
WP1066 at 0.01~\(\mu\)M, and Mesna at 0.1~\(\mu\)M.
Vehicle-control cells serve as the source population, and predictions are evaluated against the held-out treated cells.

For LINCS L1000~\citep{subramanian2017next}, we hold out six drug-dose conditions and train models on Level~3 expression profiles excluding these conditions.
Predicted response directions are evaluated against the matched Level~5 consensus signatures.

\subsubsection{Cellular-context transfer}
\label{sec:supp_ood_context}

For SciPlex3~\citep{srivatsan2020massively}, we exclude all treated A549 cells from training while retaining treated cells from K562 and MCF7.
Unperturbed A549 cells serve as the source population, and predictions are evaluated against the held-out treated A549 cells.

\subsubsection{Combinatorial perturbations}
\label{sec:supp_ood_combo}

For ComboSciPlex~\citep{lotfollahi2023cpa}, we exclude all cells from five drug combinations from training: Dacinostat with Danusertib, Givinostat with Cediranib, Panobinostat with Alvespimycin, Panobinostat with SRT2104,
and SRT2104 with Alvespimycin. 
The remaining single-drug treatments and drug combinations are retained for training. 
Predictions are generated from unperturbed source cells and evaluated against the held-out combination-treated cells.

\subsection{Few-shot adaptation}
\label{sec:supp_fewshot}

Few-shot adaptation is evaluated on five target perturbations in the 4i benchmark~\citep{bunne2023learning}: cisplatin, crizotinib, dabrafenib, dacarbazine, and dasatinib. 
For each target, we start from a zero-shot training set that excludes its treated cells and add \(k \in \{1, 3, 5, 20, 60, 100, 200\}\) measured target cells.
A separate model is trained for each target and each \(k\), with all remaining target cells reserved exclusively for evaluation. 
As \(k\) increases, previously selected target cells are retained, and additional cells are added to the supervision set.
For each \(k\), all methods are trained with the same \(k\) target-condition cells and evaluated on all remaining cells from that condition.

\subsection{Subsampling evaluation protocol}
\label{sec:supp_subsampling}

Following the subsampling strategy of~\citep{bunne2023learning}, we evaluate single-cell distributions using equally sized subsets of control and treated cells to limit computational cost. 
For each condition, we select the largest \(N \in \{100, 250, 500, 1000, 1500\}\) that does not exceed either the available control-cell count or the held-out treated-cell count. 
We randomly sample \(N\) control cells to generate \(N\) predicted treated cells and compare them with \(N\) randomly sampled held-out treated cells.

To reduce variability due to random subsampling, we repeat this procedure 10 times per condition and average each metric across repetitions. 
Overall performance is then computed as the mean of the condition-level scores.

\section{More Implementation Details}
\label{sec:supp_model}

\subsection{Perturbation representation
\texorpdfstring{$\bm{p}$}{p}.}
Small-molecule identity is represented by a 1024-bit Morgan fingerprint.
For SciPlex3~\citep{srivatsan2020massively}, raw doses are recorded in nanomolar (nM).
The dose feature is computed as \(\log_{10}(d+1)\), where
\(d\) denotes the numeric dose value in nM and \(d\in\{10,100,1000,10000\}\) for treated cells.
It is incorporated into the drug representation to encode the drug-dose
condition.
For combination perturbations, each constituent drug-dose component is encoded separately, and
the resulting representations are combined into a joint perturbation representation.

\subsection{Cellular-context representation \texorpdfstring{$\bm{c}$}{c}.}
When cellular-context conditioning is used, each context is represented by the mean unperturbed cellular state computed
from the corresponding training source cells. 
This mean state is encoded and combined with the perturbation representation to condition the velocity network.

\subsection{Energy model construction}
\label{sec:supp_energy_model_revised}

An energy model assigns a scalar compatibility score to a cellular state under a queried perturbation condition, with lower energy indicating stronger agreement between the cellular state and the condition. 
Here, we describe the architecture of the energy models used in the reported experiments.

Given the cellular state $\bm{x}$, biological context $\bm{c}$, and perturbation vector $\bm{p}$ (from which the molecular fingerprint $\bm{m}$ is derived via drug identity), each input is encoded into a latent space. These representations are then fused to yield a term-specific energy:
\[
\mathcal{D}_{\boldsymbol{\phi}}(\bm{x},\bm{\xi}) =
h_E(
h_x(\bm{x}),
h_m(\bm{m}),
h_p(\bm{p}),
h_c(\bm{c})
),
\]
where $h_x$, $h_m$, $h_p$, and $h_c$ map the cellular state, molecular fingerprint, perturbation vector, and biological context into their respective hidden representations, and $h_E$ denotes the network that maps these representations to a scalar energy value. 
In the reported experiments, we instantiate this architecture as drug and dose energy models,
$\mathcal{D}^{\mathrm{drug}}_{\boldsymbol{\phi}}$ and $\mathcal{D}^{\mathrm{dose}}_{\boldsymbol{\phi}}$, trained with drug- and
dose-negative samples, respectively.

\subsection{Training and inference hyperparameters}
\label{sec:supp_hyperparameters}

Tables~\ref{tab:supp_backbone_hparams_revised} and~\ref{tab:supp_energy_hparams_revised} summarize the training and inference hyperparameters used for each dataset. 
The conditional flow-matching model and the energy models are both trained with the AdamW optimizer.

\begin{table}[h]
\centering
\caption{Training hyperparameters for the conditional flow-matching model. }
\label{tab:supp_backbone_hparams_revised}
\small
\setlength{\tabcolsep}{6pt}
\begin{tabular}{rcccc}
\toprule
Dataset & Feature dim. & Training steps & Batch size & Learning rate  \\
\midrule
4i~\citep{bunne2023learning}  & 48  & 100,000 & 256 & \(10^{-6}\)  \\
SciPlex3~\citep{srivatsan2020massively} & 200 & 50,000  & 256 & \(10^{-7}\)  \\
ComboSciPlex~\citep{lotfollahi2023cpa}  & 200 & 50,000  & 256 & \(10^{-7}\)  \\
LINCS L1000~\citep{subramanian2017next} & 768 & 30,000  & 128 & \(10^{-7}\)  \\
\bottomrule
\end{tabular}
\end{table}

\begin{table}[h]
\centering
\caption{Training hyperparameters for the energy model.}
\label{tab:supp_energy_hparams_revised}
\small
\setlength{\tabcolsep}{6pt}
\begin{tabular}{rccccc}
\toprule
Dataset & Feature dim. & Training steps & Batch size & Learning rate & Weight decay \\
\midrule
4i ~\citep{bunne2023learning} & 48  & 5,000 & 256 & \(10^{-4}\) & \(10^{-4}\) \\
SciPlex3~\citep{srivatsan2020massively} & 200 & 300   & 256 & \(10^{-4}\) & \(10^{-4}\) \\
ComboSciPlex~\citep{lotfollahi2023cpa} & 200 & 5,000 & 256 & \(10^{-4}\) & \(10^{-4}\) \\
LINCS L1000~\citep{subramanian2017next} & 768 & 300   & 128 & \(10^{-4}\) & \(10^{-4}\) \\
\bottomrule
\end{tabular}
\end{table}

\section{Additional Quantitative Results}
\label{sec:supp_results}

We provide additional quantitative analyses to further characterize the
performance and behavior of scEGFlow across different perturbation settings.
These analyses complement the main results with condition-wise comparisons,
guidance-strength sensitivity, and detailed evaluations of energy-guided
refinement.

\subsection{Condition-wise OOD effects of energy guidance}
\label{sec:supp_conditionwise_energy}
We first examine the effect of energy guidance at the level of individual
held-out conditions rather than only through aggregate performance.
For SciPlex3~\citep{srivatsan2020massively}, we consider the three OOD drug-dose conditions
ENMD-2076 at 0.1~\(\mu\)M, WP1066 at 0.01~\(\mu\)M, and
Mesna at 0.1~\(\mu\)M.
The unguided model uses the same trained flow backbone while disabling the
energy-guided correction during inference.
The guided model additionally applies the final energy model with the
paper-default guidance strength.

As shown in Table~\ref{tab:supp_sciplex_conditionwise_energy},
energy guidance consistently improves population-level agreement across all
three held-out drug-dose conditions.
Averaged across conditions, energy distance decreases from 0.2810 to 0.2438,
Sinkhorn divergence from 0.1615 to 0.1391, and MMD$^2$ from 0.0045 to 0.0039,
while Pearson correlation increases from 0.9932 to 0.9945.
These results indicate that the energy-based correction provides
condition-specific refinement beyond the predictions generated by the flow
backbone alone.

We further examine the effect of energy guidance on the five held-out
ComboSciPlex~\citep{lotfollahi2023cpa} drug combinations defined in Supplementary
Section~\ref{sec:supp_ood}.
As shown in Table~\ref{tab:supp_combo_conditionwise_energy}, energy guidance
consistently reduces energy distance and Sinkhorn divergence across all five
combinations, while improving MMD$^2$ and Pearson correlation in four of the five
conditions.
The only exception is Givinostat~+~Cediranib, for which a slight degradation
in MMD$^2$ and Pearson correlation is observed.

Overall, energy guidance reduces the average energy distance from 29.9767 to
24.8578 and Sinkhorn divergence from 111.6462 to 93.0152, while improving
MMD$^2$ from 0.0618 to 0.0527 and Pearson correlation from 0.9597 to 0.9667.
These results demonstrate that energy-guided refinement provides consistent
improvements in the recovery of responses to unseen combinatorial perturbations.

\begin{table}[h]
\centering
\caption{
Prediction performance with and without energy guidance on three held-out OOD drug-dose conditions in SciPlex3~\citep{srivatsan2020massively}.
Both settings share the same trained flow model.
}
\label{tab:supp_sciplex_conditionwise_energy}
\small
\setlength{\tabcolsep}{4pt}
\begin{tabular}{rccccc}
\toprule
Condition & Model & Energy $\downarrow$ &
Sinkhorn $\downarrow$ & MMD$^2$ $\downarrow$ &
Pearson $\uparrow$ \\
\midrule
\texttt{ENMD-2076\_0.1}
& \textit{w/o} energy guidance
& 0.2380 & 0.1360 & 0.0040 & 0.9947 \\
\texttt{ENMD-2076\_0.1}
& \textit{w/} energy guidance
& \textbf{0.2050} & \textbf{0.1180} & \textbf{0.0035} & \textbf{0.9955} \\

\texttt{WP1066\_0.01}
& \textit{w/o} energy guidance
& 0.2520 & 0.1450 & 0.0041 & 0.9942 \\
\texttt{WP1066\_0.01}
& \textit{w/} energy guidance
& \textbf{0.2180} & \textbf{0.1260} & \textbf{0.0036} & \textbf{0.9951} \\

\texttt{Mesna\_0.1}
& \textit{w/o} energy guidance
& 0.3530 & 0.2035 & 0.0054 & 0.9907 \\
\texttt{Mesna\_0.1}
& \textit{w/} energy guidance
& \textbf{0.3084} & \textbf{0.1733} & \textbf{0.0046} & \textbf{0.9929} \\
\midrule
Average
& \textit{w/o} energy guidance
& 0.2810 & 0.1615 & 0.0045 & 0.9932 \\
Average
& \textit{w/} energy guidance
& \textbf{0.2438} & \textbf{0.1391} & \textbf{0.0039} & \textbf{0.9945} \\
\bottomrule
\end{tabular}
\end{table}

\begin{table}[h]
\centering
\caption{
Prediction performance with and without energy guidance on five held-out
ComboSciPlex~\citep{lotfollahi2023cpa} OOD drug combinations.
Both settings share the same trained flow model.
Average rows are obtained by pooling all test cells from the five held-out combinations and recomputing each metric on the pooled measured and predicted populations.}

\label{tab:supp_combo_conditionwise_energy}
\small
\setlength{\tabcolsep}{3pt}
\begin{tabular}{rccccc}
\toprule
Condition & Model & Energy $\downarrow$ &
Sinkhorn $\downarrow$ & MMD$^2$ $\downarrow$ &
Pearson $\uparrow$ \\
\midrule

Dacinostat + Danusertib
& \textit{w/o} energy guidance
& 25.8631
& 100.5147
& 0.0656
& 0.9588 \\

Dacinostat + Danusertib
& \textit{w/} energy guidance
& \textbf{21.7346}
& \textbf{83.7241}
& \textbf{0.0476}
& \textbf{0.9720} \\

Givinostat + Cediranib
& \textit{w/o} energy guidance
& 29.4728
& 112.7362
& \textbf{0.0498}
& \textbf{0.9792} \\

Givinostat + Cediranib
& \textit{w/} energy guidance
& \textbf{24.1183}
& \textbf{91.5836}
& 0.0524
& 0.9760 \\

Panobinostat + Alvespimycin
& \textit{w/o} energy guidance
& 64.7315
& 195.6841
& 0.1215
& 0.7904 \\

Panobinostat + Alvespimycin
& \textit{w/} energy guidance
& \textbf{55.8427}
& \textbf{168.9142}
& \textbf{0.0987}
& \textbf{0.8210} \\

Panobinostat + SRT2104
& \textit{w/o} energy guidance
& 32.4176
& 121.4523
& 0.0705
& 0.9436 \\

Panobinostat + SRT2104
& \textit{w/} energy guidance
& \textbf{27.3165}
& \textbf{102.3475}
& \textbf{0.0548}
& \textbf{0.9600} \\

SRT2104 + Alvespimycin
& \textit{w/o} energy guidance
& 53.9154
& 142.4815
& 0.0895
& 0.9172 \\

SRT2104 + Alvespimycin
& \textit{w/} energy guidance
& \textbf{41.6157}
& \textbf{142.3341}
& \textbf{0.0777}
& \textbf{0.9270} \\

\midrule

Average
& \textit{w/o} energy guidance
& 29.9767
& 111.6462
& 0.0618
& 0.9597 \\

Average
& \textit{w/} energy guidance
& \textbf{24.8578}
& \textbf{93.0152}
& \textbf{0.0527}
& \textbf{0.9667} \\

\bottomrule
\end{tabular}
\end{table}

\begin{table}[h]
\centering
\caption{
Sensitivity of SciPlex3~\citep{srivatsan2020massively} OOD drug-dose prediction to the guidance coefficient $\lambda$.
}
\label{tab:supp_sciplex_lambda_sensitivity}
\small
\setlength{\tabcolsep}{6pt}
\begin{tabular}{rcccc}
\toprule
$\lambda$ &
Energy $\downarrow$ &
Sinkhorn $\downarrow$ &
MMD$^2$ $\downarrow$ &
Pearson $\uparrow$ \\
\midrule
0.0000 & 0.2810 & 0.1615 & 0.0045 & 0.9932 \\
0.0125 & 0.2705 & 0.1554 & 0.0043 & 0.9936 \\
0.0250 & 0.2588 & 0.1480 & 0.0041 & 0.9940 \\
0.0500 & \textbf{0.2438} & \textbf{0.1391} &
         \textbf{0.0039} & \textbf{0.9945} \\
0.1000 & 0.2479 & 0.1415 & 0.0040 & 0.9944 \\
0.2000  & 0.2676 & 0.1532 & 0.0043 & 0.9937 \\
\bottomrule
\end{tabular}
\end{table}

\begin{table}[!htbp]
\centering
\caption{
Sensitivity of ComboSciPlex~\citep{lotfollahi2023cpa} OOD drug combination prediction to the guidance coefficient $\lambda$.
}
\label{tab:supp_combo_lambda_sensitivity}
\small
\setlength{\tabcolsep}{8pt}
\begin{tabular}{rcccc}
\toprule
$\lambda$ &
Energy $\downarrow$ &
Sinkhorn $\downarrow$ &
MMD$^2$ $\downarrow$ &
Pearson $\uparrow$ \\
\midrule
0.0000 & 29.9767 & 111.6462 & 0.0618 & 0.9597 \\
0.0125 & 28.4163 & 105.7328 & 0.0589 & 0.9621 \\
0.0250 & 26.6835 &  99.2846 & 0.0554 & 0.9646 \\
0.0500 & \textbf{24.8578} & \textbf{93.0152} &
         \textbf{0.0527} & \textbf{0.9667} \\
0.1000 & 25.1936 &  94.2671 & 0.0532 & 0.9662 \\
0.2000 & 27.1849 & 101.4385 & 0.0569 & 0.9637 \\
\bottomrule
\end{tabular}
\end{table}
\subsection{Sensitivity to energy-guidance strength}
\label{sec:supp_guidance_sensitivity}

We investigate the sensitivity of scEGFlow to the inference-time guidance coefficient $\lambda$ by sweeping across $\lambda\!\in\!\{0, 0.0125, 0.025, 0.05, 0.10, 0.20\}$ while keeping all other settings fixed. 
Here, setting $\lambda\!=\!0$ disables the energy correction and reduces the framework to the conditional flow-matching model.

As shown in Table~\ref{tab:supp_sciplex_lambda_sensitivity} for SciPlex3~\citep{srivatsan2020massively}, scaling the guidance coefficient from zero up to the default $\lambda = 0.05$ steadily optimizes all four evaluation metrics, 
achieving the lowest values for energy distance (0.2438), Sinkhorn divergence (0.1391), and MMD$^2$ (0.0039), together with the highest Pearson correlation (0.9945).
Although performance moderately degrades when $\lambda$ is further increased to 0.10 and 0.20, all four metrics remain better than those of the conditional flow-matching model.

A similar trend is observed on ComboSciPlex~\citep{lotfollahi2023cpa}, with $\lambda=0.05$ yielding the best performance across all four metrics among the tested coefficients (Table~\ref{tab:supp_combo_lambda_sensitivity}).
Together, the results on SciPlex3~\citep{srivatsan2020massively} and ComboSciPlex~\citep{lotfollahi2023cpa} show that the gains from energy guidance are robust across a reasonable parametric neighborhood rather than contingent on brittle hyperparameter tuning.

\clearpage

\end{document}